\documentclass[a4paper,10pt,numbers=noenddot,twoside=on,headinclude=true,footinclude=false,headsepline=true]{scrartcl}

\usepackage[utf8]{inputenc}
\usepackage[T1]{fontenc}
\usepackage{textcomp}
\usepackage[english]{babel}
\usepackage{color}
\usepackage{amssymb}
\usepackage{amsmath}
\usepackage{amsfonts}
\usepackage{amsopn}
\usepackage{amsbsy}
\usepackage{enumerate}
\usepackage[plainpages=false,pdfpagelabels,breaklinks=true,pdfborderstyle={/S/U/W 1.2}]{hyperref}
\usepackage{graphics}
\usepackage{units}

\numberwithin{equation}{section} 
\numberwithin{table}{section} 
\numberwithin{figure}{section} 

\usepackage[amsmath,thmmarks,thref,hyperref]{ntheorem}

\theoremstyle{plain}

\theorembodyfont{\upshape}

\theoremstyle{nonumberplain}

\theoremsymbol{\ensuremath{_\Box}}

\usepackage[style=alphabetic,
	citestyle=alphabetic,
	firstinits=true,
	backend=biber,
	hyperref=auto,
	sorting=nyt,
	maxcitenames=3,
	maxbibnames=99,
	minnames=3,
	arxiv=pdf,
	url=false,
	isbn=false,
	dateabbrev=true,
	datezeros=true,
	bibencoding=utf8]{biblatex}
\newbibmacro{string+doi}[1]{%
	\iffieldundef{doi}{#1}{\href{http://dx.doi.org/\thefield{doi}}{#1}}}
\DeclareFieldFormat
	{title}{\usebibmacro{string+doi}{\mkbibemph{#1}}\isdot}
\DeclareFieldFormat
	[article,inbook,incollection,inproceedings,patent,thesis,unpublished]
	{title}{\usebibmacro{string+doi}{\mkbibemph{#1}}\isdot}
\DeclareFieldFormat
	[article,inbook,incollection,inproceedings,patent,thesis,unpublished]
	{pages}{#1}
\DeclareFieldFormat
	[article]
	{journaltitle}{#1\isdot}
\DeclareFieldFormat
	[article]
	{volume}{\textbf{#1}}
\DefineBibliographyStrings{english}{pages={}}
\DeclareBibliographyDriver{article}{%
	\usebibmacro{bibindex}%
	\usebibmacro{begentry}%
	\usebibmacro{author/translator+others}%
	\setunit{\labelnamepunct}\newblock
	\usebibmacro{title}%
	\usebibmacro{journal+issuetitle}%
	\setunit*{\addcomma\space}
	\printfield{pages}
	\setunit*{\addcomma\space}
	\printfield{year}
	\usebibmacro{finentry}%
}

\newbibmacro*{journal+issuetitle}{%
	\usebibmacro{journal}%
	\setunit*{\addspace}%
	\iffieldundef{series}
	{}
	{\newunit
		\printfield{series}%
		\setunit{\addspace}}%
	\iffieldundef{volume}
		{}
		{\printfield{volume}%
		\setunit{\addspace}}%
	\setunit*{\addcolon\space}%
	\usebibmacro{issue}%
	\newunit}

\usepackage[scaled]{beramono}
\usepackage{helvet}
\usepackage[charter]{mathdesign} 
\SetMathAlphabet{\mathcal}{normal}{OMS}{cmsy}{m}{n} 
\SetMathAlphabet{\mathcal}{bold}{OMS}{cmsy}{m}{n} 
\providecommand{\ie}{i.~e.~}
\providecommand{\eg}{e.~g.~}
\providecommand{\cf}{cf.~}

\providecommand{\R}{\mathbb{R}}

\providecommand{\C}{\mathbb{C}}
\renewcommand{\C}{\mathbb{C}}

\providecommand{\T}{\mathbb{T}}
\renewcommand{\T}{\mathbb{T}}

\providecommand{\Z}{\mathbb{Z}}
\providecommand{\ii}{\mathrm{i}}
\providecommand{\e}{\mathrm{e}}
\renewcommand{\Re}{\mathrm{Re} \,}
\renewcommand{\Im}{\mathrm{Im} \,}

\providecommand{\eps}{\varepsilon}

\providecommand{\ran}{\mathrm{ran} \, }

\providecommand{\dd}{\mathrm{d}}
\providecommand{\id}{\mathrm{id}}

\providecommand{\order}{\mathcal{O}}

\providecommand{\abs}[1]{\left \lvert #1 \right \rvert}
\providecommand{\sabs}[1]{\lvert #1 \vert}
\providecommand{\babs}[1]{\bigl \lvert #1 \bigr \rvert}

\providecommand{\bnorm}[1]{\bigl \lVert #1 \bigr \rVert}

\providecommand{\scpro}[2]{\left \langle #1 , #2 \right \rangle}

\providecommand{\bscpro}[2]{\bigl \langle #1 , #2 \bigr \rangle}

\providecommand{\sopro}[2]{\vert #1 \rangle \langle #2 \vert}

\providecommand{\Rot}{\mathrm{Rot}}
\providecommand{\Zak}{\mathcal{Z}}

\usepackage{array}
\usepackage{slashbox}
\usepackage{multirow}
\usepackage{float}
\restylefloat{table}
\usepackage[table]{xcolor}
\definecolor{lightgray}{rgb}{0.86,0.86,0.86}

\bibliography{/Users/max/Library/texmf/tex/latex/max/bibliography}

\title{On the Role of Symmetries \\ in the Theory of Photonic Crystals}
\author{Giuseppe De Nittis \& Max Lein}

\begin{document}

\maketitle
\vspace{-9mm}
\begin{center}
	$^{\ast}$ Department Mathematik, 
	Universität Erlangen-Nürnberg \linebreak
	Cauerstrasse 11, 
	D-91058 Erlangen, 
	Germany \linebreak
	{\footnotesize \href{mailto:denittis@math.fau.de}{\texttt{denittis@math.fau.de}}}
	\medskip
	\\
	$^{\ast\ast}$ University of Toronto \& Fields Institute, 
	Department of Mathematics \linebreak
	40 St{.} George Street, 
	Toronto, ON M5S 2E4, 
	Canada \linebreak
	{\footnotesize \href{mailto:max.lein@utoronto.ca}{\texttt{max.lein@utoronto.ca}}}
\end{center}
\begin{abstract}
	We discuss the role of the symmetries in photonic crystals and classify them according to the Cartan-Altland-Zirnbauer scheme. Of particular importance are complex conjugation $C$ and time-reversal $T$, but we identify also other significant symmetries. Borrowing the jargon of the classification theory of topological insulators, we show that $C$ is a “particle-hole-type symmetry” rather than a “time-reversal symmetry” if one consider the Maxwell operator in the first-order formalism where the dynamical Maxwell equations can be rewritten as a Schrödinger equation; The symmetry which implements physical time-reversal is a “chiral-type symmetry”. We justify by an analysis of the band structure why the first-order formalism seems to be more advantageous than the second-order formalism. Moreover, based on the Schrödinger formalism, we introduce a class of effective (tight-binding) models called Maxwell-Harper operators. Some considerations about the breaking of the “particle-hole-type symmetry” in the case of gyrotropic crystals are added at the end of this paper.
\end{abstract}
{\scriptsize \textbf{PACS 2010:} 41.20.Jb, 42.70.Qs, 78.20.-e}

\newpage
\tableofcontents

\section{Introduction} 
\label{intro}
Roughly speaking, a photonic crystal (PhC) is to light what a crystalline solid is to an electron. Based on this analogy, experiments have been proposed which realize “quantum-like systems” in PhCs. On the other hand, many well-known effects from solid state physics have been anticipated in PhCs. That is how edge currents in PhCs have been predicted \cite{Raghu_Haldane:quantum_Hall_effect_photonic_crystals:2008,Ochiai_Onoda:edge_states_photonic_graphene:2009,Lu_Joannopoulos_et_al:edge_modes_3d_photonic_crystal:2012} and observed \cite{Wang_et_al:edge_modes_photonic_crystal:2008,Wang_et_al:unidirectional_backscattering_photonic_crystal:2009,Fu_et_al:blockage_edge_modes:2011,Plotnik_et_al:edge_modes_photonic_graphene:2012,Rechtsman_Plotnik_et_al:edge_states_photonic_graphene:2013}. 

However, this correspondence between electrodynamics and quantum mechanics is not one-to-one, and there are aspects where these differences become crucial. A priori there is no way of knowing when it breaks down or even if “analogous” phenomena have the same explanation. For instance, the analogy to the Bloch electron suggests that the existence of topologically protected edge states in PhCs can be explained by the \emph{bulk-edge correspondence} (proved under various levels of generality in \cite{Hatsugai:Chern_number_edge_states:1993,Kellendonk_Schulz-Baldes:quantization_edge_currents:2004}). Its validity is still an open problem, and it cannot merely be assumed but eventually needs to be established by a first-principles derivation. 

This paper focuses on the role of symmetries, because breaking or imposing the correct symmetries becomes crucial for the observation of topological effects. Our main purpose is to give a complete classification of Maxwell operators $M_w$ according to the Cartan-Altland-Zirnbauer (CAZ) scheme \cite{Altland_Zirnbauer:superconductors_symmetries:1997,Schnyder_Ryu_Furusaki_Ludwig:classification_topological_insulators:2008}. This necessitates a reformulation of the Maxwell equations as a first-order, Schrödinger-type equation. The structural similarities between Maxwell operators and massless Dirac operators are crucial for the correct identification of relevant symmetries, chief among them are complex conjugation $C$ and time-reversal $T$. Consequently, we obtain a exhaustive classification of \emph{“photonic topological insulators”} \cite{Khanikaev_et_al:photonic_topological_insulators:2013,Rechtsman_Zeuner_et_al:photonic_topological_insulators:2013,Lin_et_al:topological_photonic_states:2014}, including all expected topological invariants for each CAZ class. For instance, this allows us to predict which CAZ classes support non-trivial invariants which are expected to play a crucial role in a first-principles derivation of a photonic bulk-edge correspondence.

More specifically, our main points are: 
\begin{enumerate}[(1)]
	\item The dynamical Maxwell equations can be recast in the form of the “Schrödinger equation”~\eqref{description:eqn:Schroedinger_Maxwell}, a first-order equation in time. This Schrödinger-type point of view allows one to adapt many techniques initially developed for quantum systems to problems of classical electromagnetism, \eg space-adiabatic perturbation theory \cite{PST:effective_dynamics_Bloch:2003,DeNittis_Lein:sapt_photonic_crystals:2013} and the classification of symmetries. While it is true that complex conjugation $C$ leaves the \emph{second-order} formulation of the Maxwell equations~\eqref{first_vs_second_order:eqn:Maxwell_wave} invariant, it is \emph{not} a time-reversal symmetry. $C$ does not implement time-reversal and in the parlance of classification theory of topological insulators it acts as a \emph{“particle-hole symmetry”}. The well-known physical time-reversal $T : (\mathbf{E},\mathbf{H}) \mapsto (\mathbf{E},-\mathbf{H})$ classifies as \emph{“chiral symmetry”}; Note that in the second-order formalism, $T$ becomes a trivial symmetry. \item Since the Schrödinger equation is a \emph{first}-order equation in time, it is the \emph{first}-order classification of $C$ as “particle-hole symmetry” and $T$ as a “chiral symmetry” which matters for the subsequent analysis. A correct CAZ classification of symmetries is impossible in the second-order formalism: when applying symmetries to the \emph{square} of the Maxwell operator $M_w^2$ one can no longer distinguish “particle-hole” and “time-reversal symmetries” as well as “chiral symmetries” and linear, commuting symmetries from one another. For PhCs with real material weights the Maxwell operator (\cf equation~\eqref{Max_Schroedinger:eqn:Maxwell_operator}) is of symmetry class D, DIII or BDI rather than AI, so \emph{even in non-gyrotropic media}, Chern numbers (or other topological obstructions) associated to single, isolated bands need not be zero!
	\item Nevertheless, we show the absence of topological effects in PhCs with real weights for an important class of initial conditions, namely \emph{real} fields: the presence of the $C$-symmetry implies that real initial states emerge as linear combinations of conjugate pairs of Bloch functions. Thanks to the $C$-symmetry, the total Chern number associated with a pair of conjugate states is zero because the Chern numbers of symmetrically related bands are equal in magnitude but have opposite sign. However, we do not know whether these arguments necessarily imply the absence of \emph{all} topological effects: Maxwell operators with $C$-symmetry have other $\Z$- or $\Z_2$-valued topological invariants which may be non-zero. Further research is needed.
	\item The study of effective dynamics for real initial states in photonic cyrstals becomes a \emph{bona fide multiband problem} since single bands can never support real states. That is particularly significant for approaches which have derived effective ray optics equations, because effective single band equations do not describe the evolution of real states. Deriving multiband ray optics for real initial states is still an open problem; here, the main obstacle is that real states are, to use a term from quantum mechanics, entangled, and one needs to control intraband terms (\cf the discussion in \cite[Section~5]{DeNittis_Lein:sapt_photonic_crystals:2013}).
	\item A derivation of ray optics equations in the standard second-order framework is made more difficult because one is no longer able to distinguish genuine band crossings from “artificial” ones (compare Figures~\ref{Max_Schroedinger:fig:band_spectrum} and \ref{first_vs_second_order:fig:abs_band_spectrum}). This is because in the second-order formulation outgoing ($\omega_n(k) > 0$) and incoming ($\omega_n(k) < 0$) frequency bands cannot be distinguished, and the presence of any chiral or particle-hole symmetry lead to symmetries of the form $\omega_n(k) \leftrightarrow -\omega_n(\pm k)$. Thus, there are no isolated, non-degenerate bands in the $\abs{\omega}$ band spectrum of the most common photonic crystals with $C$- or $T$-symmetry. Moreover, it is $\omega_n$ rather than $\sabs{\omega_n}$ which enters the ray optics equations. 
	\item The CAZ scheme classifies operators in terms of one unitary and/or one antiunitary operator as well as their product. For the Maxwell operator, $C$ and $T$ are not the only choices, and we systematically explore alternate symmetries. In particular, we enumerate the conditions placed on the material weights by the presence of symmetries. Here, it is crucial that we work in the first-order Schrödinger-type framework to identify the nature of each of these symmetries properly. The structural similarity of the Maxwell operator and massless Dirac operators helps to find an exhaustive list of symmetries which are relevant for the purpose of CAZ classification. Hence, we obtain a complete classification of photonic topological insulators (see Table~\ref{symmetries:table:all_symmetry_classes}) and we tabulate the topological invariants for each class (see Table~\ref{symmetry:table:K_groups}). At least 5 of the 10 CAZ classes have already been considered in the physics literature.
	\item We propose the \emph{Maxwell-Harper operator}~\eqref{Maxwell_Harper:eqn:Maxwell_Harper} for a conjugate pair of bands as a simple model operator for non-gyrotropic PhCs in analogy to the usual Harper operator. It is a $2 \times 2$ matrix operator and exists even if the bands carry non-zero Chern charge. 
\end{enumerate}
In what follows, we will derive and expound on these assertions. 

The paper is roughly structured as follows: we will first reformulate the Maxwell equations as a Schrödinger-type equation in Section~\ref{Max_Schroedinger}, derive the frequency band spectrum, expound on the significance of complex conjugation and introduce the proper time-reversal operation. Then we will juxtapose first- and second-order formalism in Section~\ref{first_vs_second_order}. The Cartan-Altland-Zirnbauer classification of Maxwell operators is the topic of Section~\ref{symmetry}. Here, we will explain the nature of $C$ and $T$ in the CAZ framework, explore other symmetries, discuss the CAZ classification of Maxwell operators and finish with a discussion of topological invariants. The Maxwell-Harper operator is introduced in Section~\ref{Maxwell_Harper}. We close the paper by a discussion of \emph{gyrotropic} materials where the electric permittivity $\eps$ and the magnetic permeability $\mu$ are hermitian matrix-valued functions with non-zero imaginary parts.

\section{The first-order Schrödinger formalism} 
\label{Max_Schroedinger}
The claim that light and a quantum particles behave similarly is fundamentally a statement about their \emph{dynamics}. So one way to make such a claim rigorous is by reformulating the dynamical Maxwell equations as a Schrödinger-type equation. This first-order Schrödinger-type formalism of electromagnetism allows one to adapt tools developed for analyzing quantum problems; for instance, the authors were able to derive effective dynamics for adiabatically perturbed PhCs \cite{DeNittis_Lein:sapt_photonic_crystals:2013} by adapting a technique initially developed for adiabatic quantum systems \cite{PST:effective_dynamics_Bloch:2003,DeNittis_Lein:Bloch_electron:2009}. 

The propagation of electromagnetic waves in a linear, three-dimensional medium are governed by the two dynamical Maxwell equations, 
\begin{subequations}\label{Max_Schroedinger:eqn:Maxwell_dynamical}
	\begin{align}
		\eps \, \partial_t \mathbf{E}(t,x) &= \nabla \times \mathbf{H}(t,x) 
		, 
		\\
		\mu \, \partial_t \mathbf{H}(t,x) &= - \nabla \times \mathbf{E}(t,x) 
		, 
	\end{align}
\end{subequations}
whereas the absence of sources is described by 
\begin{subequations}\label{Max_Schroedinger:eqn:Maxwell_sources}
	\begin{align}
		\nabla \cdot \eps \mathbf{E}(t,x) &= 0 
		, 
		\\
		\nabla \cdot \mu \mathbf{H}(t,x) &= 0 
		. 
	\end{align}
\end{subequations}
The properties of the material enter through the electric permittivity tensor $\eps$ and the magnetic permeability tensor $\mu$. While most materials are non-gyrotropic, \ie $\eps(x)$ and $\mu(x)$ are \emph{real-symmetric}, there are cases when $\eps(x)$ and $\mu(x)$ are hermitian with non-zero imaginary part (see \eg \cite{Yeh_Chao_Lin:Faraday_effect:1999,Wu_Levy_Fratello_Merzlikin:gyrotropic_photonic_crystals:2010,Kriegler_Rill_Linden_Wegener:bianisotropic_photonic_metamaterials:2010,Esposito_Gerace:photonic_crystals_broken_TR_symmetry:2013}). Throughout this paper, we assume that $\eps$ and $\mu$ are positive, bounded with bounded inverse. To simplify the notation we shall refer to $w = (\eps,\mu)$ as \emph{material weights} and use $\overline{w} = (\bar{\eps},\bar{\mu})$ to denote the complex conjugate weights. 

To the best of our knowledge, the idea to express the dynamical Maxwell equations as a Schrödinger equation 
\begin{align}
	\ii \partial_t \Psi &= M_w \Psi
	\label{description:eqn:Schroedinger_Maxwell}
\end{align}
originated in a paper by Birman and Solomyak \cite{Birman_Solomyak:L2_theory_Maxwell_operator:1987}. Here, the electromagnetic field $\Psi = (\mathbf{E},\mathbf{H})$ plays the role of the wave function and the \emph{Maxwell operator}
\begin{align}
	M_w = \left (
	\begin{matrix}
		0 & + \ii \, \eps^{-1} \, \nabla^{\times} \\
		- \ii \, \mu^{-1} \, \nabla^{\times} & 0 \\
	\end{matrix}
	\right )
	\label{Max_Schroedinger:eqn:Maxwell_operator}
\end{align}
takes the place of the quantum Schrödinger operator $H = \frac{1}{2m} (-\ii \hbar \nabla)^2 + V$. Throughout the paper we use the short-hand $v^{\times} \mathbf{E} = v \times \mathbf{E}$, \eg $\nabla^{\times} \mathbf{E} = \nabla \times \mathbf{E}$ denotes the curl. The material weights $w = (\eps,\mu)$ also enter into the definition of the scalar product 
\begin{align}
	\bscpro{\Psi}{\Psi'}_w &
	= \bscpro{\mathbf{E}}{\eps \mathbf{E}'} + \bscpro{\mathbf{H}}{\mu \mathbf{H}'}
	\notag \\
	&= \int_{\R^3} \dd x \, \Bigl ( \mathbf{E}(x) \cdot \eps(x) \, \mathbf{E}'(x) + \mathbf{H}(x) \cdot \mu(x) \, \mathbf{H}'(x) \Bigr )
	, 
	\label{Max_Schroedinger:eqn:weighted_scalar_product}
\end{align}
and the corresponding Hilbert space $L^2_w(\R^3,\C^6)$ is $L^2(\R^3,\C^6)$ equipped with the weighted scalar product $\scpro{\, \cdot \,}{\cdot \,}_w$. Note that complex conjugation is contained in $v \cdot w := \sum_{j = 1}^3 \overline{v_j} \, w_j$. 

This weighted scalar product provides a decomposition of electromagnetic waves into longitudinal and transversal component: a quick computation shows that gradient fields are $\scpro{\, \cdot \,}{\cdot \,}_w$-orthogonal to fields satisfying~\eqref{Max_Schroedinger:eqn:Maxwell_dynamical} \cite[Section~3]{DeNittis_Lein:adiabatic_periodic_Maxwell_PsiDO:2013}. Moreover, the Maxwell operator is hermitian, $\scpro{\Psi}{M_w \Psi'}_w = \scpro{M_w \Psi}{\Psi'}_w$, and consequently, the time-evolution $\e^{- \ii t M_w}$ is unitary. This leads to the conservation of field energy 
\begin{align}
	\mathcal{E}(\mathbf{E},\mathbf{H}) 
	= \frac{1}{2} \int_{\R^3} \dd x \, \Bigl ( \mathbf{E}(x) \cdot \eps(x) \, \mathbf{E}(x) + \mathbf{H}(x) \cdot \mu(x) \, \mathbf{H}(x) \Bigr )
	= \frac{1}{2} \, \bnorm{(\mathbf{E},\mathbf{H})}_w^2
	. 
	\label{Max_Schroedinger:eqn:field_energy}
\end{align}

\subsection{The frequency band picture} 
\label{Max_Schroedinger:band_picture}
The particularity of photonic crystals is that the material weights $(\eps,\mu)$ are periodic with respect to some lattice $\Gamma = \mathrm{span}_{\Z} \{ e_1 , e_2 , e_3 \}$ spanned by three (non-unique) fundamental vectors. Now one proceeds as if $M_w$ were a periodic Schrödinger operator: we employ the Bloch-Floquet-Zak transform \cite{Zak:dynamics_Bloch_electrons:1968,Kuchment:Floquet_theory:1993} 
\begin{align}
	(\Zak \Psi)(k,y) &= \sum_{\gamma \in \Gamma} \e^{- \ii k \cdot (y + \gamma)} \, \Psi(y + \gamma)
	\label{Max_Schroedinger:eqn:Zak_transform}
\end{align}
to change representation ($\Zak$ maps onto the space-periodic part of Bloch functions). As explained in \cite[Section~3]{DeNittis_Lein:adiabatic_periodic_Maxwell_PsiDO:2013} the operator $M_w$ is unitarily equivalent to a family of Maxwell operators 
\begin{align}
	M_w(k) &= W \, \Rot
	= \left (
	\begin{matrix}
		\eps^{-1} & 0 \\
		0 & \mu^{-1} \\
	\end{matrix}
	\right ) \, 
	\left (
	\begin{matrix}
		0 & - (-\ii \nabla + k)^{\times} \\
		+ (-\ii \nabla + k)^{\times} & 0 \\
	\end{matrix}
	\right )
	\label{Max_Schroedinger:eqn:fiber_Maxwell}
\end{align}
depending on crystal momentum $k$ where $M_w(k)$ acts on $\Gamma$-periodic electromagnetic fields $\psi = (\psi^E,\psi^H)$. This gives rise to Bloch functions $\varphi_n(k)$ and Bloch frequency bands $\omega_n(k)$,%
\begin{figure}[t]
	\centering
		\resizebox{100mm}{!}{\includegraphics{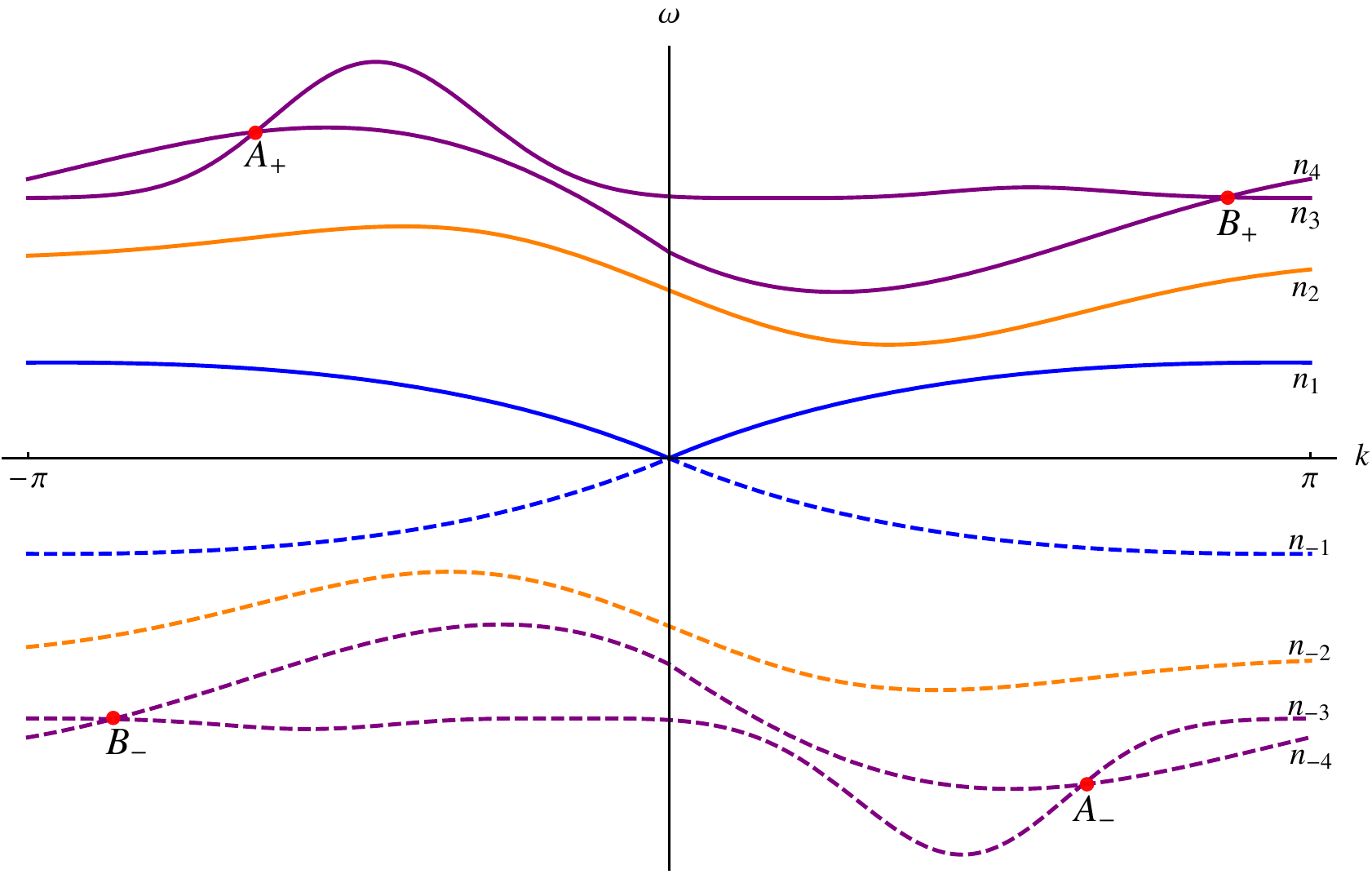}}
	\caption{A sketch of a typical band spectrum of $M_w(k)$ for a non-gyrotropic photonic crystal with broken time-reversal symmetry (\ie $\eps$, $\mu$ and $\chi$ from equation~\eqref{symmetry:eqn:W_with_chi} are real and non-zero). The $2+2$ ground state bands ($\pm n_1$) with linear dispersion around $k = 0$ are blue. Positive frequency bands are drawn using solid lines while the lines for the symmetrically-related negative frequency bands are in the same color, but dashed. }
	\label{Max_Schroedinger:fig:band_spectrum}
\end{figure}
\begin{align}
	M_w(k) \varphi_n(k) &= \omega_n(k) \, \varphi_n(k) 
	. 
	\label{Max_Schroedinger:eqn:eigenvalue_equation}
\end{align}
$M_w(k)$ has a flat band $\omega(k) = 0$ due to unphysical gradient fields; Bloch functions associated to non-zero frequency bands are automatically source-free \cite[Section~3]{DeNittis_Lein:adiabatic_periodic_Maxwell_PsiDO:2013}, 
\begin{align*}
	(\nabla + \ii k) \cdot \eps \varphi_n^E &= 0 
	,
	\\
	(\nabla + \ii k) \cdot \mu \varphi_n^H &= 0 
	. 
\end{align*}
Schematically, the frequency band spectrum looks as in Figure~\ref{Max_Schroedinger:fig:band_spectrum}: there are 2 “ground state bands” with approximately linear dispersion at $k \approx 0$. Apart from $k = 0$, bands do not touch $\omega = 0$. Note that there are bands of positive and negative frequency; the signs correspond to outgoing and incoming complex Bloch waves. This can already be inferred from \eqref{Max_Schroedinger:eqn:Maxwell_operator}, the Maxwell operator looks very similar to a massless Dirac operator
\begin{align*}
	M_w = - W \, \sigma_2 \otimes \nabla^{\times}
\end{align*}
where $W$ is given by equation~\eqref{symmetry:eqn:W_with_chi}. Note that the figure depicts the band spectrum of a \emph{non}-gyrotropic PhCs with broken time-reversal symmetry, leading to the point symmetry in the spectrum; in the presence of time-reversal symmetry, all band functions would in addition be even. With the exception of these symmetry considerations the band spectra of gyrotropic PhCs share all other features. 

\subsection{Complex conjugation as a particle-hole symmetry} 
\label{Max_Schroedinger:PH_symmetry}
Complex conjugation $C \Psi(x) := \overline{\Psi(x)}$ induces a relation between $M_w$ and the Maxwell operator $M_{\overline{w}}$ with complex conjugate material weights, 
\begin{align}
	C \, M_w \, C = - M_{\overline{w}} 
	.
	\label{Max_Schroedinger:eqn:PH_symmetry}
\end{align}
Consequently, Maxwell operators $M_w$ for non-gyrotropic media ($\overline{w} = w$) can be of class D, DIII or BDI in the Cartan-Altland-Zirnbauer (CAZ) classification scheme depending on the presence of additional symmetries \cite{Altland_Zirnbauer:superconductors_symmetries:1997,Schnyder_Ryu_Furusaki_Ludwig:classification_topological_insulators:2008}. In other words $C$ acts as a \emph{particle hole symmetry} which squares to $+\id$. However, $C$ \emph{cannot} be interpreted as implementing time-reversal, because unlike for the Schrödinger evolution group we in fact obtain\footnote{Note that $C$ is not an anti-unitary map between $L^2_w(\R^3,\C^6)$ and itself but only as a map $L^2_w(\R^3,\C^6)$ and the Hilbert space with conjugate weights $L^2_{\overline{w}}(\R^3,\C^6)$.} 
\begin{align}
	C \, \e^{- \ii t M_w} \, C &= \e^{+ \ii t \, C M_w C}
	= \e^{- \ii t M_{\overline{w}}} 
	. 
	\label{Max_Schroedinger:eqn:C_commutes_evolution_group}
\end{align}
Instead, complex conjugation interchanges incoming and outgoing (complex) Bloch waves; This was first noted in \cite[Section~III]{Bergmann:gyrotropic_Maxwell:1982} for purely homogeneous media where the material weights are constant, but was not linked explicitly to complex conjugation. 

Translating equation~\eqref{Max_Schroedinger:eqn:PH_symmetry} to the Bloch-Floquet-Zak representation involves the symmetry $C^{\Zak} = \Zak \, C \, \Zak^{-1}$,  
\begin{align}
	\bigl ( C^{\Zak} \varphi \bigr )(k,y) = \overline{\varphi(-k,y)}
	, 
	\label{Max_Schroedinger:eqn:C_Zak}
\end{align}
and leads to 
\begin{align}
	C \, M_w(k) \, C &= - M_{\overline{w}}(-k) 
	. 
	\label{Max_Schroedinger:eqn:PH_symmetry_M}
\end{align}
In other words, we have $M_w(k) \varphi_n(k) = \omega_n(k) \, \varphi_n(k)$ if and only if 
\begin{align*}
	M_{\overline{w}}(k) \, \bigl ( C^{\Zak} \varphi_n \bigr )(k) &= - \omega_n(-k) \, \bigl ( C^{\Zak} \varphi_n \bigr )(k)
	. 
\end{align*}
Equation~\eqref{Max_Schroedinger:eqn:PH_symmetry_M} explains the point symmetry in the band spectrum of non-gyrotropic materials where $\overline{w} = w$: \emph{frequency bands come in pairs} $\bigl ( \omega_n(k) , - \omega_n(-k) \bigr )$ with Bloch functions $\bigl ( \varphi_n(k) \, , \, (C^{\Zak} \varphi_n)(k) \bigr )$ (dashed and solid lines of the same color in Figure~\ref{Max_Schroedinger:fig:band_spectrum}). These Bloch waves are necessarily \emph{complex}, because $\varphi_n$ and $C^{\Zak} \varphi_n$ are eigenfunctions to frequency bands of different sign.\footnote{Even for ground state bands, \ie the bands with linear dispersion around $k = 0$, a real Bloch basis can only be chosen at $k = 0$.} 

\subsection{Implementation of time-reversal symmetry} 
\label{Max_Schroedinger:time-reversal}
On physical grounds it is misleading to call $C$ a “time-reversal symmetry”, though: in view of equation~\eqref{Max_Schroedinger:eqn:C_commutes_evolution_group} time-reversal 
\begin{align}
	\bigl ( \mathbf{E}(t) , \mathbf{H}(t) \bigr ) \mapsto \bigl ( \mathbf{E}(-t) , - \mathbf{H}(-t) \bigr )
	\label{Max_Schroedinger:eqn:time-reversal_condition}
\end{align}
is not implemented by $C$ as in the case of quantum mechanics. Instead, the correct operation 
\begin{align*}
	T = \sigma_3 \otimes \id : (\mathbf{E},\mathbf{H}) \mapsto (\mathbf{E},-\mathbf{H})
\end{align*}
is \emph{linear} (as opposed to anti-linear) and flips the sign of the magnetic field strength. In the CAZ classification $T$ is a “chiral-type symmetry” \cite{Schnyder_Ryu_Furusaki_Ludwig:classification_topological_insulators:2008}; we emphasize that both, in case of $C$ and $T$ the names from the CAZ classification do \emph{not} correspond to their physical meaning in electromagnetism (\cf Table~\ref{symmetry:table:meaning_symmetries}). A straightforward computation shows $T \, M_w \, T = - M_w$, and consequently the linearity of $T$ implies 
\begin{align}
	T \, \e^{- \ii t M_w} \, T = \e^{+ \ii t M_w}
	. 
	\label{Max_Schroedinger:eqn:time-reversal}
\end{align}
This equation, however, is just another way of saying that $T$ implements \eqref{Max_Schroedinger:eqn:time-reversal_condition}, 
\begin{align*}
	\left (
	\begin{matrix}
		\mathbf{E}(t) \\
		(-\mathbf{H})(t) \\
	\end{matrix}
	\right ) &= 
	\e^{- \ii t M_w} \, \left (
	\begin{matrix}
		\mathbf{E} \\
		- \mathbf{H} \\
	\end{matrix}
	\right )
	= \e^{- \ii t M_w} \, T \left (
	\begin{matrix}
		\mathbf{E} \\
		\mathbf{H} \\
	\end{matrix}
	\right )
	= T \, \e^{+ \ii t M_w} \left (
	\begin{matrix}
		\mathbf{E} \\
		\mathbf{H} \\
	\end{matrix}
	\right )
	= \left (
	\begin{matrix}
		\mathbf{E}(-t) \\
		- \mathbf{H}(-t) \\
	\end{matrix}
	\right )
	.
\end{align*}
Also $T$ gives rise to a symmetry in the band spectrum: taking into account that $T$ is linear, time-reversal yields the fiber-wise relation 
\begin{align*}
	T \, M_w(k) \, T = - M_w(k)
	. 
\end{align*}
Consequently, also the $T$-symmetry pairs frequency bands $\bigl ( \omega_n(k) , - \omega_n(k) \bigr )$ with Bloch functions $\bigl ( \varphi_n(k) , T \varphi_n(k) \bigr )$. 

Let us briefly mention that in \cite[eqns.~(18a)--(18d)]{Kong:symmetries_bianisotropic_media:1972} $J = T \, C$ has been proposed as time-reversal symmetry for Maxwell equations with complex material weights. The difference between $T$ and $J$ only becomes significant for gyrotropic photonic crystals where $w \neq \overline{w}$, because then instead of \eqref{Max_Schroedinger:eqn:time-reversal} the operator $J$ intertwines the evolution of $M_w$ and $M_{\overline{w}}$, 
\begin{align*}
	J \, \e^{- \ii t M_w} \, J = e^{+ \ii t M_{\overline{w}}}
	. 
\end{align*}
Seeing as $T$ satisfies equation~\eqref{Max_Schroedinger:eqn:time-reversal} and agrees with the way time-reversal is defined in other literature (see \eg \cite[Table~6.1]{Jackson:electrodynamics:1998} or \cite[Chapter~7]{Altman_Suchy:reciprocity_time-reversal_electromagnetics:2011}), we will continue to refer to $T$ as time-reversal. We will pick up the topic of symmetries in Section~\ref{symmetry}. 

\subsection{Real states in non-gyrotropic photonic crystals} 
\label{Max_Schroedinger:physical_states}
Very often the initial states of interest are 
\emph{real} electromagnetic waves $(\mathbf{E},\mathbf{H}) = C (\mathbf{E},\mathbf{H})$. By definition Zak transforms of such functions are “real” with respect to $C^{\Zak} = \Zak \, C \, \Zak^{-1}$, \ie 
\begin{align*}
	C (\mathbf{E},\mathbf{H}) = (\mathbf{E},\mathbf{H}) 
	\; \; \Leftrightarrow \; \; 
	C^{\Zak} \, \Zak (\mathbf{E},\mathbf{H}) = \Zak (\mathbf{E},\mathbf{H}) 
	. 
\end{align*}
The action of $C^{\Zak}$ given by \eqref{Max_Schroedinger:eqn:C_Zak} is no longer just pointwise complex conjugation, and to avoid confusion we call functions $\psi(k)$ in Bloch-Floquet-Zak representation \emph{Real} if $(C^{\Zak} \psi)(k) = \overline{\psi(-k)} = \psi(k)$. This is also consistent with the terminology used in the theory of Real vector bundles \cite{Atiyah:K_theory_reality:1966,DeNittis_Gomi:AI_bundles:2014}.

Let us focus on states that are localized in a narrow frequency range, \ie states associated to a family of Bloch bands which do not cross or merge with other bands. In the simplest case, we only need to consider a single, non-degenerate bands $\omega_+ > 0$ and its symmetric twin $\omega_-(k) = - \omega_+(-k)$ whose Bloch functions $\varphi_-(k) = \bigl ( C^{\Zak} \varphi_+ \bigr )(k) = \overline{\varphi_+(-k)}$ are related by complex conjugation. Then the two Real solutions 
\begin{align*}
	\psi_{\Re}(k) &= \tfrac{1}{\sqrt{2}} \, \bigl ( \varphi_+(k) + \varphi_-(k) \bigr )
	\\
	\psi_{\Im}(k) &= \tfrac{1}{\ii \, \sqrt{2}} \, \bigl ( \varphi_+(k) - \varphi_-(k) \bigr )
\end{align*}
are Real and Imaginary part of $\varphi_+$; if we introduce the Real part operator $\Re^{\Zak} = \tfrac{1}{2} \bigl ( 1 + C^{\Zak} \bigr )$ and the Imaginary part operator $\Im^{\Zak} = \tfrac{1}{\ii \, 2} \bigl ( 1 - C^{\Zak} \bigr )$, then we can succinctly write $\psi_{\Re} = \sqrt{2} \, \Re^{\Zak} \varphi_+$ and $\psi_{\Im} = \sqrt{2} \, \Im^{\Zak} \varphi_+$. Real states associated to the band $\omega_+$ are \emph{real} linear combinations of $\psi_{\Re}$ and $\psi_{\Im}$, and these real linear combinations always depend on both, $\varphi_+$ \emph{and} $\varphi_-$. Hence, finding effective dynamics for real states is a \emph{multi}-band problem, one which is still unsolved. 

\subsection{Two- vs{.} three-dimensional PhCs} 
\label{Max_Schroedinger:2d_vs_3d}
All of our arguments generalize to two-dimensional PhCs as the $2d$ Maxwell operator is a restriction of \eqref{Max_Schroedinger:eqn:Maxwell_operator}. Thus, also in two dimensions $C$ is a particle-hole symmetry and frequency bands on the two-dimensional Brillouin zone come in conjugate pairs. 

To make our arguments self-contained, let us sketch a derivation (see also \cite[Chapter~7.2.5]{Kuchment:math_photonic_crystals:2001} for the isotropic case): suppose the material weights $w = (\eps,\mu)$ are of both of the form 
\begin{align}
	w &= \left (
	\begin{matrix}
		w_1 & u + \ii \, v & 0 \\
		u + \ii \, v & w_2 & 0 \\
		0 & 0 & w_3 \\
	\end{matrix}
	\right )
	= \left (
	\begin{matrix}
		\tilde{w} & 0 \\
		0 & w_3 \\
	\end{matrix}
	\right )
	\label{Max_Schroedinger:eqn:2d_weights}
\end{align}
\ie they factor into two blocks. As a consequence the fields in the $x_1 x_2$-plane and along the $x_3$-direction are orthogonal, \eg 
\begin{align*}
	\scpro{(E_1,E_2,0)}{\eps (0,0,E_3)} = 0 
	, 
\end{align*}
and similarly for the magnetic field. Moreover, electric fields of the form $(E_1,E_2,0)$ drive the dynamics for the magnetic field only along the $x_3$-direction, $\partial_t \mathbf{H} = - \mu^{-1} \, \nabla \times \mathbf{E} = \bigl ( 0,0,\partial_t H_3 \bigr )$. Hence, if we start with a \emph{transverse electric} (TE) mode $\bigl ( (E_1,E_2,0) ,  (0 , 0 , H_3) \bigr )$, the time-evolved state will be of the same form. One can repeat the same arguments for \emph{transverse magnetic} (TM) modes $\bigl ( (0,0,E_3) , (H_1,H_2,0) \bigr )$.

Now let us impose a second assumption on the material weights, namely that they are independent of $x_3$. Then we can make a product ansatz $\Psi(x_1,x_2,x_3) = \Phi(x_1,x_2) \, \e^{+ \ii k_3 \, x_3}$ for the electromagnetic fields where the component depending on $x_3$ is just a plane wave. The $2d$ Maxwell operator emerges after choosing $k_3 = 0$ (meaning the fields are independent of $x_3$), \ie 
\begin{align*}
	M_{w,2d} = \left (
	\begin{matrix}
		0 & + \ii \, \eps^{-1} \, (\partial_1,\partial_2,0)^{\times} \\
		- \ii \, \mu^{-1} \, (\partial_1,\partial_2,0)^{\times} & 0 \\
	\end{matrix}
	\right )
\end{align*}
where $(\partial_1,\partial_2,0)^{\times} \mathbf{E} = (\partial_1,\partial_2,0) \times \mathbf{E}$. Electromagnetic fields of finite energy are now elements of $L^2_w(\R^2,\C^6)$ with a weighted scalar product defined analogously to \eqref{Max_Schroedinger:eqn:weighted_scalar_product}. 

The block structure of the material weights leads to a block decomposition of the Maxwell operator $M_{w,2d} = M_{\mathrm{TE}} \oplus M_{\mathrm{TM}}$ induced by splitting electromagnetic fields into TE and TM modes, 
\begin{align*}
	L^2_w(\R^2,\C^6) = L^2_{\mathrm{TE}} \oplus L^2_{\mathrm{TM}}
	. 
\end{align*}
These two operators can be compactly written as $3 \times 3$-matrix-valued operators, \eg  
\begin{align*}
	M_{\mathrm{TE}} &= \left (
	\begin{matrix}
		0 & 0 & 
		\multirow{2}{*}{$+ \ii \, \tilde{\eps}^{-1} \left (
		\begin{matrix}
			+ \partial_2 \\
			- \partial_1 \\ 
		\end{matrix}
		\right )$}
		 \\
		0 & 0 &  \\
		+ \ii \, \mu_3^{-1} \, \partial_2 & - \ii \, \mu_3^{-1} \, \partial_1 & 0 \\
	\end{matrix}
	\right )
\end{align*}
is the form of the Maxwell operator for TE modes $(E_1,E_2,H_3)$. This block structure also means that TE and TM components are independent, \eg one can compute their (two-dimensional) frequency bands and Bloch functions separately. Given that we have derived the $2d$ Maxwell operator from the full, three-dimensional Maxwell equations, $M_{w,2d}$ inherits properties such as the particle hole-type symmetry. Consequently, also $2d$ frequency bands and Bloch functions come as conjugate pairs. And while the particle hole-type symmetry does not force single bands to be topologically trivial (the Chern number associated to $\varphi_n$ need not be zero), real and imaginary part of $\varphi_n$ are. From the viewpoint of topological insulators this is not surprising: the $2d$ Maxwell operator is of class D in the Altland-Zirnbauer classification scheme, so one expects to find a $\Z$-valued topological index. 

\section{First- vs{.} second-order formalism} 
\label{first_vs_second_order}
Most of the time, at least implicitly, the second-order equation 
\begin{align}
	\bigl ( \partial_t^2 + M_w^2 \bigr ) \Psi = 0 
	\label{first_vs_second_order:eqn:Maxwell_wave}
\end{align}
is considered instead of \eqref{Max_Schroedinger:eqn:eigenvalue_equation}. From a practical point of view, this has a number of advantages, most importantly, electric and magnetic components decouple and one obtains two second-order PDEs. And only one of the two equations needs to be solved. Moreover, in two dimensions this leads to two \emph{scalar} equations, one for the TM and another for the TE modes. These simplifications allow for a more efficient treatment. Clearly, for non-gyrotropic materials where $M_w = M_{\overline{w}}$, complex conjugation leaves \eqref{first_vs_second_order:eqn:Maxwell_wave} invariant. 

The eigenvalue problem that is usually solved in other works reads 
\begin{align*}
	M_w(k)^2 \varphi_n(k) = \bigl ( \lambda_n(k) \bigr )^2 \, \varphi_n(k) 
\end{align*}
where $\lambda_n(k) = \sabs{\omega_n(k)}$ is taken to be positive. This eigenvalue problem is subtly different from \eqref{Max_Schroedinger:eqn:eigenvalue_equation}, because the information whether the Bloch wave is outgoing ($\omega_n > 0$) or incoming ($\omega_n < 0$) is discarded. 

The information contained in the sign of $\omega_n$ is critical when one wants to reconstruct solutions to the \emph{dynamical} problem. 
The similarity of \eqref{first_vs_second_order:eqn:Maxwell_wave} to the wave equation suggests to rewrite it as 
\begin{align*}
	 \bigl ( \partial_t + \ii M_w(k) \bigr ) \, \bigl ( \partial_t - \ii M_w(k) \bigr ) \Psi 
	 = 0 
\end{align*}
in Bloch-Floquet-Zak representation, and we see that any solution of the second-order equation has to be a linear combination of an outgoing and an incoming wave. 

\begin{figure}[t]
	\centering
		\resizebox{100mm}{!}{\includegraphics{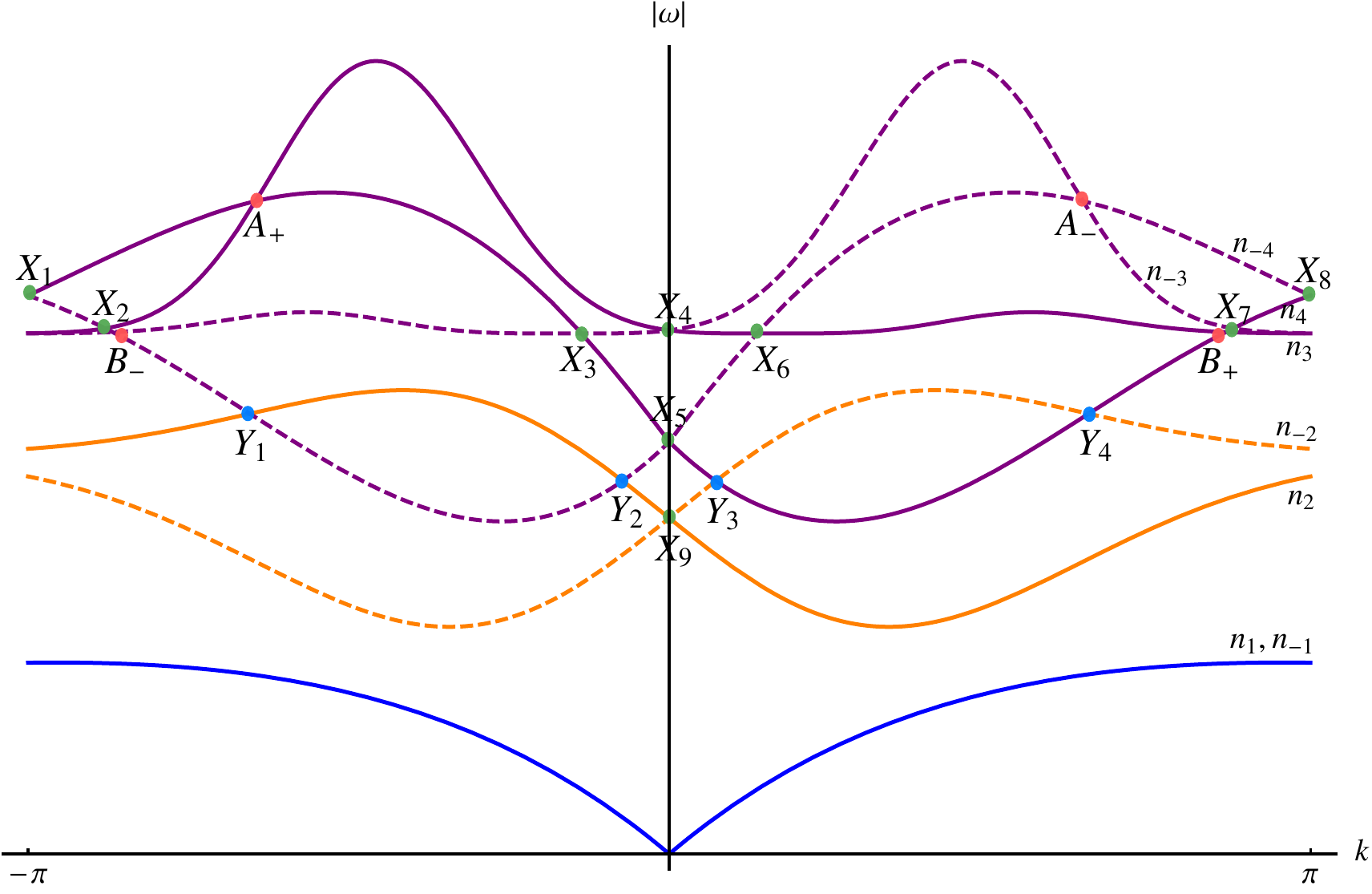}}
	\caption{The absolute value of the band spectrum from Figure~\ref{Max_Schroedinger:fig:band_spectrum}. Aside from proper band intersections (points labeled $A_j$ and $B_j$), numerous artificial intersections are introduced ($X_j$ and $Y_j$). $X_j$ denotes intersections between symmetrically related bands (bands of the same color) while the $Y_j$ are fictitious intersections between unrelated bands. Note that one cannot tell from the $\abs{\omega}$ spectrum whether time-reversal symmetry is present; if it is, all frequency bands are of even degeneracy. }
	\label{first_vs_second_order:fig:abs_band_spectrum}
\end{figure}

But even if one is not interested in the dynamical problem, the loss of information by discarding the sign of $\omega_n$ also affects the analysis of the frequency band spectrum, topological quantities and effective models. Primarily there are two types of situations which are the starting point for further research, namely (i) isolated bands and (ii) band crossings (especially conical crossings). Or put another way, what matters are the locations of band crossings and degeneracies. If we take the absolute value of the spectrum of our fictitious Maxwell operator from Figure~\ref{Max_Schroedinger:fig:band_spectrum}, the much more convoluted frequency band picture of Figure~\ref{first_vs_second_order:fig:abs_band_spectrum} emerges. Compared to the signed band spectrum, many \emph{artificial} band crossings appear (the points $X_j$ and $Y_j$). These artificial crossings will have no interesting physical effects associated to them because these bands are in fact decoupled from one another (\eg bands $n_{\pm 2}$ intersects with bands $n_{\pm 4}$ in Figure~\ref{first_vs_second_order:fig:abs_band_spectrum}). 

The presence of symmetries such as $C$ or $T$ necessarily generates degeneracies in the $\abs{\omega}$ spectrum, because each of these symmetries lead to symmetric pairings of bands. While in PhCs with $C$-symmetry the pairing $\omega_n(k) \leftrightarrow - \omega_n(-k)$ leads to a degeneracy only at $k = 0$ in the $\abs{\omega}$ band picture, frequency bands of PhCs with $T$-symmetry have even degeneracy everywhere, because of $\omega_n(k) \leftrightarrow - \omega_n(k)$. Put another way, if the PhC has $C$- or $T$-symmetry, then \emph{there are no isolated non-degenerate bands in the second-order formalism} -- even if they are isolated in the first-order formalism. In case symmetries are absent and there is no relation between the positive and negative frequency spectrum, but nevertheless folding up the negative-frequency part still creates artificial band crossings.

The absence of isolated bands in the second-order framework for the most common classes of PhCs, \ie those with $C$- or $T$-symmetry, also makes a derivation of correct ray optics equations more difficult. 
Indeed, recovering the sign of the frequency band $\omega_n(k)$ is not just important for checking whether $\omega_n(k)$ is indeed isolated and all band crossings are artificial. The sign of $\omega_n$ is crucial when solving ray optics equations such as \cite[eqns.~(42)--(43)]{Raghu_Haldane:quantum_Hall_effect_photonic_crystals:2008} or \cite[eqn.~(45)]{DeNittis_Lein:sapt_photonic_crystals:2013} (see also \cite{Onoda_Murakami_Nagaosa:geometrics_optical_wave-packets:2006,Esposito_Gerace:photonic_crystals_broken_TR_symmetry:2013}): in case $\omega_n(k) < 0$ the ray optics flows associated to $\omega_n(k)$ and $\sabs{\omega_n(k)}$ will be qualitatively different from one another. 

A distinction between artificial and proper band crossings also enters when one wants to discuss conical intersection and avoided conical crossings using, say, graphene-type models \cite{Onoda_Murakami_Nagaosa:topological_nature_polarization:2004,Raghu_Haldane:quantum_Hall_effect_photonic_crystals:2008,Ochiai_Onoda:edge_states_photonic_graphene:2009,DeNittis_Lein:piezo_graphene:2013}. These graphene-type models encapsulate two interesting features: first of all, there is a link between symmetry breaking and the opening of a gap, and secondly, the existence of topologically non-trivial phases \cite{Hasegawa_Kohmoto:quantum_Hall_effect_graphene:2006,Ochiai_Onoda:edge_states_photonic_graphene:2009}. Artificial band crossings will behave very differently (the points $X_j$ and $Y_j$) from proper band crossings (the points $A_j$ and $B_j$) under symmetry breaking: artificial band crossings will persist while for proper ones gaps may open. 

Even if the spectral information can somehow be reconstructed, labeling bands in this fashion \emph{artificially alters the topology of the Bloch bands}. Given that topological terms also enter into the \emph{dynamical} equations, \eg the Berry curvature appears in the ray optics equations, labeling bands properly is crucial. Moreover, we note that one needs the magnetic \emph{and} electric component in order to compute Chern numbers; in the single-band case where $c_1 = \frac{1}{2\pi} \int_{\mathbb{B}} \dd k \, \Omega_n(k)$ the right-hand side is guaranteed to be an integer only if one takes $\Omega_n(k)$ to be the rotation of 
\begin{align*}
	\mathcal{A}_n(k) = \int \dd y \, \varphi_n^E(k,y) \cdot \eps(y) \varphi_n^E(k,y) + \int \dd y \, \varphi_n^H(k,y) \cdot \mu(y) \varphi_n^H(k,y)
\end{align*}
rather than the rotation of the first or second term of $\mathcal{A}_n$ only. 

Finally, a correct identification of the nature of symmetries for the purpose of CAZ classification is not possible in the second-order framework, because the action of many symmetries becomes trivial when considering $M_w^2$ instead. More on that below. 
%

\section{The CAZ classification of Maxwell operators} 
\label{symmetry}
The first-order Schrödinger-type formalism presented in Section~\ref{Max_Schroedinger} allows one to systematically adapt tools developed for analyzing topological phases of quantum systems. Concretely, we apply the Cartan-Altland-Zirnbauer (CAZ) classification scheme \cite{Altland_Zirnbauer:superconductors_symmetries:1997,Schnyder_Ryu_Furusaki_Ludwig:classification_topological_insulators:2008} to PhCs.
Similarly to crystalline solids, PhCs can be classified using the CAZ classification scheme; this gives rise to different classes of “photonic topological insulators” \cite{Khanikaev_et_al:photonic_topological_insulators:2013,Rechtsman_Zeuner_et_al:photonic_topological_insulators:2013,Lin_et_al:topological_photonic_states:2014} which are characterized by different topological invariants. 
The aim of this section is to identify for the first time the CAZ classes of PhCs and their topological invariants.

To include more general linear and lossless media, we now consider Maxwell operators $M_w = W \, \Rot$ with material weights of the form 
\begin{align}
	W^{-1} = \left (
	\begin{matrix}
		\eps & \chi \\
		\chi^* & \mu \\
	\end{matrix}
	\right ) 
	\label{symmetry:eqn:W_with_chi}
\end{align}
such that $W^{-1}$ is bounded, invertible and has a bounded inverse $W$. The assumption that the material is lossless is equivalent to $W = W^*$. Material weights of this form have been discussed in the physics literature (see \eg \cite{Padilla:group_theory_metamaterials:2007,Lin_et_al:topological_photonic_states:2014}). 

The starting point of a CAZ classification is the choice of one or two symmetry operators, one unitary and/or one antiunitary which have to square to $\pm \id$. These are symmetries of the free Maxwell operator $\Rot$, and the question arises for which material weights these symmetries persist?

\subsection{Classification via $C$ and $T$} 
\label{symmetry:classification}
Our discussion in Sections~\ref{Max_Schroedinger:PH_symmetry}--\ref{Max_Schroedinger:time-reversal} suggests to use the antiunitary $C$ and the unitary $T$; we will also need their product $J = T \, C$. All of these operators square to $+ \id$ and are indeed symmetries of the free Maxwell operator, 
\begin{align*}
	C \, \Rot \, C &= - \Rot 
	, 
	\\
	T \, \Rot \, T &= - \Rot 
	, 
	\\
	J \, \Rot \, J &= + \Rot 
	. 
\end{align*}
Because the CAZ scheme was initially developed for quantum problems, the CAZ designations of $C$ as a “particle-hole symmetry”, of $T$ as a “chiral symmetry” and of $J$ as a “time-reversal symmetry” do not match the terminology of electromagnetism (\cf Table~\ref{symmetry:table:meaning_symmetries}). 
\begin{table}[t]
	\begin{tabular}{c | c | c | c | c}
		\textbf{Symmetry} & \textbf{Linear/antilinear} & \textbf{Parity} & \textbf{CAZ designation} & \textbf{Physical meaning} \\ \hline
		$C$ & antilinear & $+$ & particle-hole symmetry & real fields remain real \\
		$T$ & linear & $+$ & chiral symmetry & time-reversal \\
		$J = T C$ & antilinear & $+$ & time-reversal symmetry &  \\
	\end{tabular}
	\caption{The CAZ classification holds for arbitrary operators, but the names are derived from the quantum world. Hence, the CAZ monikers of $T$, $C$ and $J$ do not match with their physical interpretations. }
	\label{symmetry:table:meaning_symmetries}
\end{table}

Seeing as the Maxwell operator is a product of $W$ and $\Rot$, the presence or absence of symmetries solely depends on the material weights, \ie whether 
\begin{align*}
	U \, W \, U^* = + W 
\end{align*}
where $U$ stands for $C$, $T$ or $J$. (We will discuss the case where symmetry operators \emph{anti}\-commute with $W$ below.)

Because $U$ is either unitary or antiunitary, $U \, W \, U^* = \pm W$ is in fact equivalent to $U \, W^{-1} \, U^* = \pm W^{-1}$, and we will give conditions on the constituents of $W^{-1}$, namely $\eps$, $\mu$ and $\chi$, which derive from the presence of a symmetry. A series of simple, back-of-the-envelope computations then gives rise to Table~\ref{symmetry:table:symmetry_classes}; note that $\chi = \C$, for instance, is short-hand for (i) $\chi \neq 0$ and (ii) $\Im \chi \neq 0$ for otherwise the Maxwell operator is in a different symmetry class. This way, 5 out of 10 CAZ classes can be realized with PhCs; At least 4 have been considered in physics literature. 
\begin{table}[t]
	\centering
	\begin{tabular}{c | c | c | c | c }
		\parbox[c]{1.7cm}{\centering\textbf{Symmetries present}} &  \parbox[c]{1.7cm}{\centering\textbf{CAZ class}} &  \parbox[c]{0.6cm}{\centering$\eps$, $\mu$} & \parbox[c]{0.6cm}{\centering$\chi$} & \parbox[c]{1.4cm}{\centering\textbf{Realized?}} 
		\\
		 & & & & \\ \hline
		\emph{none} & A & $\C$ & $\C$ & \cite{Lin_et_al:topological_photonic_states:2014} \\
		$J$ & AI & $\R$ & $\ii \R$ & \cite{Khanikaev_et_al:photonic_topological_insulators:2013} \\
		$T$ & AIII & $\C$ & $0$ & \cite{Raghu_Haldane:quantum_Hall_effect_photonic_crystals:2008,Lu_Fu_Joannopoulos_Soljacic:Weyl_points_gyroid_PhCs:2013,Esposito_Gerace:photonic_crystals_broken_TR_symmetry:2013} \\
		$C$ & D & $\R$ & $\R$ & unknown \\
		$T$, $C$ & BDI & $\R$ & $0$ & \cite{Ochiai_Onoda:edge_states_photonic_graphene:2009} \\
	\end{tabular}
	\caption{Cartan-Altland-Zirnbauer (CAZ) classification of Maxwell operators using $C$, $T$ and $J = T \, C$. The columns labeled $\eps$, $\mu$ and $\chi$ indicate whether these matrix-valued functions are complex, real, purely imaginary or zero; note that $\chi = \ii \R$, for instance, also implies $\chi \neq 0$. }
	\label{symmetry:table:symmetry_classes}
\end{table}
%

\subsection{Extended classification and other symmetries} 
\label{symmetries:alternative_classification}
Due to its Dirac form the free Maxwell operator $\Rot = - \sigma_2 \otimes \nabla^{\times}$ has many more symmetries than just $C$ and $T$, so the question arises whether $C$, $T$ and $J = T \, C$ are the only \emph{physically relevant} operators for a classification of Maxwell operators. Three ways to extend the classification scheme come to mind: 
\begin{enumerate}[(1)]
	\item Instead of requiring $U \, W \, U^* = + W$ for some discrete symmetry $U$, we can ask for $U \, W \, U^* = - W$. 
	\item Replace $T = T_3 = \sigma_3 \otimes \id$ with $T_j = \sigma_j \otimes \id$ for $j = 1 , 2$. Here, $T_j \, \Rot \, T_j = - \Rot$ for $j = 1 , 3$ and $T_2 \, \Rot \, T_2 = + \Rot$. Similarly, we set $J_j = T_j \, C$ for the third symmetry. Note that 
	$J_2$ is the only odd symmetry, $J_2^2 = - \id$, while all others are even. 
	\item One can include parity $(P \Psi)(x) = \Psi(-x)$ as an even symmetry. 
\end{enumerate}
Even though mathematically speaking, these symmetry conditions are equally valid, it is not clear whether they are equally \emph{mathematically} or \emph{physically significant}. 

From a mathematical perspective \emph{linear} symmetry operators which \emph{commute} with the Maxwell operator are irrelevant for the CAZ classification. For instance, if $T_2 = \sigma_2 \otimes \id$ commutes with $W$, then also $T_2 \, M_w \, T_2^* = + M_w$ holds. Moreover, \emph{linear} symmetries $U$ need to intertwine Maxwell operators for the \emph{same} value of $k$, \ie $U \, M_w(k) \, U^* = \pm M_w(k)$, whereas \emph{antilinear} symmetries $V$ have to \emph{flip the sign}, $V \, M_w(k) \, V^* = \pm M_w(-k)$. That is also why parity does not play a role in the CAZ classification: even though $P$ is linear, we have $P \, \Rot(k) \, P = - \Rot(-k)$ for the free Maxwell operator. 

We have worked out all different combinations of symmetries for the different choices of sign in Appendix~\ref{appendix:tabulated_symmetries} and arranged the different realizations of the symmetry classes in Table~\ref{symmetries:table:all_symmetry_classes}. To get a flavor let us work out the consequences of two alternate symmetry conditions: 
\newpage

\begin{table}[H]
	\newcolumntype{Q}{>{\centering\arraybackslash\footnotesize} m{2.5cm} }
	\renewcommand{\arraystretch}{1.5}
	\centering
	\begin{tabular}{c | Q | Q | Q }
		\backslashbox{\footnotesize \textbf{CAZ}}{\footnotesize \textbf{realized}} & {\normalsize 1}                                                                                                                                                             & {\normalsize 2}                                                                                                                                                                       & {\normalsize 3}                                                                                                                                                 \\ \hline 
		\parbox[c][1.1cm][c]{0.5cm}{\centering A}                                  & \cellcolor{lightgray}\parbox[c]{2.5cm}{\centering \emph{none} \\[0.5mm] {\footnotesize $w_1,w_2,w_2 , w_3 \in \C$} }                                                        &                                                                                                                                                                                       &                                                                                                                                                                 \\ \hline
		\parbox[c][1.3cm][c]{0.3cm}{\centering AIII}                               & \parbox[c]{2.5cm}{\centering $T_1 \equiv \chi$ \\[0.5mm] {\footnotesize$w_0 , w_1 \in \C$ \\ $w_2 , w_3 = 0$}}                                                              & \parbox[c]{2.5cm}{\centering $T_2 \equiv \chi$ \\[0.5mm] {\footnotesize $w_1 , w_3 \in \C$ \\ $w_0 , w_2 = 0$}}                                                                       & \cellcolor{lightgray}\parbox[c]{2.5cm}{\centering          $T_3 \equiv \chi$ \\[0.5mm] {\footnotesize $w_0 , w_3 \in \C$ \\ $w_1 , w_2 = 0$}}                   \\ \hline \hline
		\parbox[c][1.3cm][c]{0.3cm}{\centering AI}                                 & \parbox[c]{2.5cm}{\centering $J_1 \equiv + \mathrm{TR}$ \\[0.5mm] {\footnotesize$w_0 , w_1 , w_2 \in \R$ \\ $w_3 \in \ii \R$}}                                              & \cellcolor{lightgray}\parbox[c]{2.5cm}{\centering $J_3 \equiv + \mathrm{TR}$ \\[0.5mm] {\footnotesize $w_0 , w_2 , w_3 \in \R$ \\ $w_1 \in \ii \R$}}                                  & \parbox[c]{2.5cm}{\centering   $C \equiv + \mathrm{TR}$ \\[0.5mm] {\footnotesize $w_0 , w_1 , w_3 \in \ii \R$ \\ $w_2 \in \R$}}                                 \\ \hline
		\parbox[c][1.3cm][c]{0.3cm}{\centering AII}                                & \parbox[c]{2.5cm}{\centering $J_2 \equiv - \mathrm{TR}$ \\[0.5mm] {\footnotesize$w_0 \in \ii\R$ \\ $w_1 , w_2 , w_3 \in \R$}}                                               &                                                                                                                                                                                       &                                                                                                                                                                 \\ \hline \hline
		\parbox[c][1.3cm][c]{0.3cm}{\centering D}                                  & \parbox[c]{2.5cm}{\centering $J_1 \equiv + \mathrm{PH}$ \\[0.5mm] {\footnotesize$w_0 , w_1 , w_2 \in \ii \R$ \\ $w_3 \in \R$}}                                              & \parbox[c]{2.5cm}{\centering $J_3 \equiv + \mathrm{PH}$ \\[0.5mm] {\footnotesize $w_0 , w_2 , w_3 \in \ii \R$ \\ $w_1 \in \R$}}                                                       & \parbox[c]{2.5cm}{\centering   $C \equiv + \mathrm{PH}$ \\[0.5mm] {\footnotesize $w_0 , w_1 , w_3 \in \R$ \\ $w_2 \in \ii \R$}}                                 \\ \hline
		\parbox[c][1.3cm][c]{0.3cm}{\centering C}                                  & \parbox[c]{2.5cm}{\centering $J_2 \equiv - \mathrm{PH}$ \\[0.5mm] {\footnotesize$w_0 \in \R$ \\ $w_1 , w_2 , w_3 \in \ii \R$}}                                              &                                                                                                                                                                                       &                                                                                                                                                                 \\ \hline \hline
		\parbox[c][1.7cm][c]{0.3cm}{\centering BDI}                                & \parbox[c]{2.5cm}{\centering $J_1 \equiv + \mathrm{TR}$ \\ $C \equiv + \mathrm{PH}$ \\[0.5mm] {\footnotesize$w_0 , w_1 \in \R$ \\ $w_2 , w_3 = 0$}}                         & \parbox[c]{2.5cm}{\centering   $C \equiv + \mathrm{TR}$ \\ $J_1 \equiv + \mathrm{PH}$ \\[0.5mm] {\footnotesize $w_0 , w_1 \in \ii\R$ \\ $w_2 , w_3 = 0$}}                             & \parbox[c]{2.5cm}{\centering $J_3 \equiv + \mathrm{TR}$ \\   $C \equiv + \mathrm{PH}$ \\[0.5mm] {\footnotesize $w_0 , w_3 \in \R$ \\ $w_1 , w_2 = 0$}}          \\ \hline
		\parbox[c][1.7cm][c]{0.3cm}{\centering BDI}                                & \cellcolor{lightgray}\parbox[c]{2.5cm}{\centering $C \equiv + \mathrm{TR}$ \\ $J_3 \equiv + \mathrm{PH}$ \\[0.5mm] {\footnotesize$w_0 , w_3 \in \ii\R$ \\ $w_1 , w_2 = 0$}} & \parbox[c]{2.5cm}{\centering $J_3 \equiv + \mathrm{TR}$ \\ $J_1 \equiv + \mathrm{PH}$ \\[0.5mm] {\footnotesize $w_1 \in \ii \R, w_3 \in \R$ \\ $w_0 , w_2 = 0$}}                      & \parbox[c]{2.5cm}{\centering $J_1 \equiv + \mathrm{TR}$ \\ $J_3 \equiv + \mathrm{PH}$ \\[0.5mm] {\footnotesize $w_1 \in\R, w_3 \in \ii \R$ \\ $w_0 , w_2 = 0$}} \\ \hline
		\parbox[c][1.7cm][c]{0.3cm}{\centering DIII}                               & \parbox[c]{2.5cm}{\centering $J_2 \equiv - \mathrm{TR}$ \\ $J_1 \equiv + \mathrm{PH}$ \\[0.5mm] {\footnotesize$w_0 \in \ii\R, w_3 \in \R$ \\ $ w_1 , w_2 = 0$}}             & \parbox[c]{2.5cm}{\centering $J_2 \equiv - \mathrm{TR}$ \\ $J_3 \equiv + \mathrm{PH}$ \\[0.5mm] {\footnotesize $w_0 \in \ii\R, w_1 \in \R$ \\ $ w_2 , w_3 = 0$}}                      & \parbox[c]{2.5cm}{\centering $J_2 \equiv - \mathrm{TR}$ \\   $C \equiv + \mathrm{PH}$ \\[0.5mm] {\footnotesize $w_1, w_3 \in \R$ \\ $w_0 , w_2 = 0$}}           \\ \hline
		\parbox[c][1.7cm][c]{0.3cm}{\centering CI}                                 & \parbox[c]{2.5cm}{\centering $J_1 \equiv + \mathrm{TR}$ \\ $J_2 \equiv - \mathrm{PH}$ \\[0.5mm] {\footnotesize$w_0 \in \R,w_3 \in \ii \R$ \\ $w_1 , w_2 = 0$}}              & \cellcolor{lightgray}\parbox[c]{2.5cm}{\centering $J_3 \equiv + \mathrm{TR}$ \\ $J_2 \equiv - \mathrm{PH}$ \\[0.5mm] {\footnotesize $w_0 \in \R, w_1 \in \ii \R$ \\ $w_2 , w_3 = 0$}} & \parbox[c]{2.5cm}{\centering   $C \equiv + \mathrm{TR}$ \\ $J_2 \equiv - \mathrm{PH}$ \\[0.5mm] {\footnotesize $w_1, w_3 \in \ii \R$ \\ $w_0 , w_2 = 0$}}       \\
	\end{tabular}
	\caption{9 of the 10 CAZ classes can be theoretically realized while the 5 shaded cases have been realized in experiment (\cf \cite{Marques_et_al:bianisotropy_negative_permeability:2002,Khanikaev_et_al:photonic_topological_insulators:2013} for class CI and the references in Table~\ref{symmetry:table:symmetry_classes}). $U \equiv \chi$ means that $U$ acts as a \emph{chiral symmetry} (linear, $U \, M_w \, U = - M_w$). TR and PH are short for \emph{time-reversal} (antilinear, $U \, M_w \, U = + M_w$) and \emph{particle-hole symmetry} (antilinear , $U \, M_w \, U = - M_w$) which are either even ($+$) or odd~($-$). We also tabulate the associated conditions on the coefficients $w_0 , \ldots , w_3$, \cf equation~\eqref{appendix:tabulated_symmetries:eqn:W_decomposition}.}
	\label{symmetries:table:all_symmetry_classes}
\end{table}

\paragraph{$T_1 \, W \, T_1 = + W$} 
\label{symmetries:alternative_classification:T_1_C}
Here, we have replaced $T_3$ by $T_1$ but kept the sign, and a quick computation shows 
\begin{align*}
	T_1 \, W \, T_1 = + W 
	\; \; \Leftrightarrow \; \; 
	\eps = \mu 
	, \; 
	\chi = \chi^*
	. 
\end{align*}
Such a symmetry is not of purely academic interest as a PhC for microwaves with yttrium-iron-garnet (YIG) rods subjected to a magnetic field in a square lattice geometry realizes complex $\eps = \mu$ and $\chi = 0$ \cite{Pozar:microwave_engineering:1998,Wang_et_al:edge_modes_photonic_crystal:2008}. In addition, this system has time-reversal symmetry ($W$ commutes with $T = T_3$). However, the additional $T_1$-symmetry does not change the CAZ class: both, $T_1$ and $T_3$ are linear, so only one of them is relevant for the CAZ classification. And in both cases, the corresponding Maxwell operator $M_w$ is in CAZ class AIII (\cf Table~\ref{symmetries:table:all_symmetry_classes}). 

Several other works \cite{Marques_et_al:bianisotropy_negative_permeability:2002,Khanikaev_et_al:photonic_topological_insulators:2013} also consider PhCs made up of split ring resonators with $\eps = \mu = \Re \eps$ and $\chi = \Im \chi = \chi^*$. The associated Maxwell operator $M_w$ is then of class CI: in addition to the symmetry $J_3$ ($\eps$, $\mu$ real and $\chi = \Im \chi$), also $T_1$ ($\eps = \mu$, $\chi = \chi^*$) and $J_2 \propto T_1 \, J_3$ are present. Hence, we identify $T_1$ as a chiral symmetry, $J_3$ as an even time-reversal symmetry and $J_2$ as an odd particle-hole symmetry, and we read off CAZ~class CI from Table~\ref{symmetries:table:all_symmetry_classes}. 

\paragraph{$C \, W \, C = - W$} 
\label{symmetries:alternative_classification:C_minus}
If one wants $C$ to act as a “time-reversal symmetry” in CAZ parlance, then 
\begin{align*}
	C \, W \, C = - W
	\; \; \Leftrightarrow \; \; 
	\overline{\eps} = - \eps 
	, \; 
	\overline{\mu} = - \mu 
	, \; 
	\overline{\chi} = - \chi 
	, 
\end{align*}
is one way to ensure $C \, M_w \, C = + M_w$. However, combining the fact that the constituents of $W$ are purely imaginary with $W^* = W$ yields that the diagonal entries of the tensors $\eps^{-1}$ and $\mu^{-1}$ vanish identically. Apart from an \emph{approximate} realization by making the purely imaginary offdiagonal entries of $W$ much larger compared to its diagonal entries, we reckon that this symmetry seems to be of purely mathematical interest. Typically the ratio of imaginary and real part of the elements of $\eps$ and $\mu$ is very small ($\sim 10^{-3}$), and even in the YIG PhC mentioned above \cite{Pozar:microwave_engineering:1998,Wang_et_al:edge_modes_photonic_crystal:2008} it is of order $\sim 1$. 

\medskip
\noindent
Obviously, there are other combinations of symmetries and signs, and while some of them may apply to certain physically-realizable PhCs, it is clear that many just yield physically unrealistic conditions on $\eps$, $\mu$ and $\chi$. 

\subsection{Topological triviality of frequency bands and Chern numbers} 
\label{symmetry:topological_triviality}
We now explore the link between the CAZ classification and topological photonic insulator that are characterized by topological invariants. More specifically, the CAZ class of a Maxwell operator determines which topological invariants can -- and which cannot -- arise. 

\newcolumntype{C}[1]{>{\centering\arraybackslash}p{#1}}
\begin{table}[ht]
	\centering
	\begin{tabular}{c | c | C{1.4cm} | C{1.4cm} | C{1.4cm} | C{1.4cm} }
		\parbox[c]{1.7cm}{\centering\textbf{Symmetries present}} & \parbox[c]{1.7cm}{\bfseries\centering CAZ class}
		& \multicolumn{4}{c}{\textbf{Reduced $K$-group in dimension}}
		\\
		 & & $d = 1$ & $d = 2$ & $d = 3$ & $d = 4$ \\ \hline
		\emph{none}  & A & $0$ & $\Z$ & $(\Z^3)$ & $\Z\oplus(\Z^6)$ \\
		$J \equiv + \mathrm{TR}$          & AI & $0$ & $0$ & $0$ & $\Z$ \\
		$T \equiv \chi$          & AIII & $\Z$ & $(\Z^2)$ & $\Z\oplus(\Z^3)$ & ($\Z^8$) \\
		$C \equiv + \mathrm{PH}$          & D & $\Z_2$ & $(\Z_2^2) \oplus \Z$ & ($\Z_2^3 \oplus \Z^3$) & ($\Z_2^4 \oplus \Z^6$) \\
		$T \equiv \chi$, $C \equiv + \mathrm{PH}$     & BDI & $\Z$ & $(\Z^2)$ & $(\Z^3)$ &  $(\Z^4)$ \\
		$J_2 \equiv - \mathrm{PH}$, $J_3 \equiv + \mathrm{TR}$ & CI & $0$ & $0$ & $\Z$ &  $(\Z^4)$ \\
	\end{tabular}
	\caption{CAZ classification of Maxwell operators using $C$, $T$ and $J = T \, C$ as well as $J_2$ and $J_3$; all of these have been considered in the literature. We have used the terminology of Table~\ref{symmetries:table:all_symmetry_classes} to indicate the nature of the symmetries in the first column. The columns labeled $d = 1 , 2 , 3 , 4$ are the reduced $K$-groups for the torus that have been computed by standard techniques \cite{Kitaev:periodic_table_topological_insulators:2009,DeNittis_Gomi:AI_bundles:2014,DeNittis_Gomi:AII_bundles:2014}. The contributions in parentheses are \emph{weak} invariants while the origin of those without brackets, the \emph{strong} invariants, can be traced back to the reduced $K$-groups over the sphere $\mathbb{S}^d$. Note that while $d = 1$ is irrelevant for the Maxwell operator, it is helpful to include $d = 1$ to identify patterns in the dimension-dependence of the reduced $K$-groups for the Brillouin torus $\T^d$. The column $d = 4$ is relevant for Maxwell operators that depend on time in a periodic fashion. }
	\label{symmetry:table:K_groups}
\end{table}

To frame the discussion, let us sketch the argument why the first Chern class, the best-known example of topological invariants, vanish for quantum Hamiltonians with time-reversal symmetry. And then we will show why this argument \emph{fails} for Maxwell operators. 

Suppose we are given a contiguous family of energy bands $\sigma_{\mathrm{rel}}(k) = \bigcup_{n \in \mathcal{I}} \{ E_n(k) \}$ indexed by $\mathcal{I} = \{ n_{\min} , \ldots , n_{\max} \}$ that is separated from all others by a gap, \eg the yellow ($\pm n_2$) or the violet bands ($\pm n_3$) in Figure~\ref{Max_Schroedinger:fig:band_spectrum}. Then the Chern numbers are associated to the family of Bloch functions or, equivalently, to the projection onto the corresponding eigenspaces, 
\begin{align}
	\pi(k) = \sum_{n \in \mathcal{I}} \sopro{\varphi_n(k)}{\varphi_n(k)}
	= \frac{\ii}{2 \pi} \int_{\gamma(k)} \dd z \, \bigl ( H(k) - z \bigr )^{-1} 
	, 
	\label{symmetry:eqn:projection}
\end{align}
which can either be expressed as a sum of rank-$1$ projections or as a Cauchy integral where the contour $\gamma(k)$ encloses only $\sigma_{\mathrm{rel}}(k)$. 

Now if there exists an antiunitary operator $C$ for which 
\begin{align}
	C \, \pi(k) \, C = \pi(-k)
	\label{symmetry:eqn:symmetry_projection}
\end{align}
holds, then it is well-known that the associated first Chern class vanishes [cite]. In the Schrödinger case where $C$ is complex conjugation and $C \, H(k) \, C = H(-k)$, equation~\eqref{symmetry:eqn:symmetry_projection} holds true: conjugating the projection with $C$ yields 
\begin{align}
	C \, \pi(k) \, C &= - \frac{\ii}{2 \pi} \int_{\overline{\gamma(k)}} \dd z \, \bigl ( C \, H(k) \, C - \bar{z} \bigr )^{-1} 
	\notag \\
	&= \frac{\ii}{2 \pi} \int_{\gamma(k)} \dd z \, \bigl ( H(-k) - z \bigr )^{-1} 
	= \pi(-k)
	, 
	\label{symmetry:eqn:symmetry_projection_H_k}
\end{align}
because $C \, H(k) \, C = H(-k)$ implies $\sigma_{\mathrm{rel}}(-k) = \sigma_{\mathrm{rel}}(k)$ is enclosed by $\gamma(k)$. Hence, the first Chern numbers associated to $\sigma_{\mathrm{rel}}(k)$ vanish. 

Despite claims in the physics literature, this argument does not carry over to the Maxwell operator, because the right-hand side of $C \, M_w(k) \, C = - M_w(-k)$ contains an additional minus sign if the material weights are real ($M_w$ is of CAZ class D or BDI). Here, we see that the first-order Schrödinger-type formalism of electrodynamics and a correct identification of the nature of $C$ is crucial. Misidentifying complex conjugation as a “time-reversal symmetry” is false, both on physical grounds and in the context of the CAZ classification (see Table~\ref{symmetry:table:meaning_symmetries}). 

However, a quick peak at Table~\ref{symmetry:table:symmetry_classes} reveals that for Maxwell operators of class AI and BDI 
\begin{align*}
	J \, M_w(k) \, J = + M_w(-k)
\end{align*}
is satisfied, and consequently, the above argument for Schrödinger operators holds verbatim after replacing $C$ with the correct “time-reversal symmetry” $J$ and $H(k)$ with $M_w(k)$. 

There is another class of Maxwell operators, where topological effects due to non-zero Chern classes are also absent, namely those of class D ($C$-symmetry present but $T$-symmetry broken). This is for a less obvious reason: here, frequency bands come in pairs, and one always needs to take symmetrically related pairs in order to be able to form Real initial states (\cf discussion in Section~\ref{Max_Schroedinger:physical_states}). So let us pick a contiguous, separated family of positive frequency bands $\sigma_+(k) = \bigcup_{n \in \mathcal{I}} \bigl \{ \omega_n(k) \bigr \}$, and define the collection of symmetrically related bands $\sigma_-(k) = \bigcup_{n \in \mathcal{I}} \bigl \{ - \omega_n(-k) \bigr \}$ as well as the associated projections $\pi_{\pm}(k)$. Then instead of \eqref{symmetry:eqn:symmetry_projection} a modification of the arguments in equation~\eqref{symmetry:eqn:symmetry_projection_H_k} yields 
\begin{align}
	C \, \pi_+(k) \, C = \pi_-(-k) 
	, 
	\label{symmetry:eqn:symmetry_projection_M_k}
\end{align}
from which we deduce that the Chern numbers of $\pi_{\pm}$ are equal in magnitude, but are of opposite sign \cite[Remark~4]{DeNittis_Lein:sapt_photonic_crystals:2013}. However, there are a few projections which do have trivial Chern numbers, for instance, \eqref{symmetry:eqn:symmetry_projection} holds for $\pi(k) = \pi_+(k) + \pi_-(k)$. Also the projections $\sopro{\psi_{\Re}(k)}{\psi_{\Re}(k)}$ and $\sopro{\psi_{\Im}(k)}{\psi_{\Im}(k)}$ onto $\Re^{\Zak} \varphi_n$ and $\Im^{\Zak} \varphi_n$ satisfy \eqref{symmetry:eqn:symmetry_projection} by construction, and thus, \emph{Chern numbers associated to real states vanish.} This explains the absence of topological effects in non-gyrotropic PhCs. 
\medskip

\noindent
Chern numbers are but one example of topological invariants, and depending on the CAZ class there may be others. To see that, we note that any projection $\pi(k)$ of the form \eqref{symmetry:eqn:projection} defines a vector bundle, the so-called \emph{Bloch bundle} (see \eg \cite[Section~]{DeNittis_Lein:exponentially_loc_Wannier:2011}), whose structure can be analyzed with standard tools of the trade such as $K$-theory. For each CAZ class it is the \emph{reduced $K$-group} which identifies the form of the topological invariants; the reduced $K$-groups for the CAZ classification from Section~\ref{symmetry:classification} has been tabulated in Table~\ref{symmetry:table:K_groups}. 

Note that $K$-theory gives no clue how to \emph{compute} topological invariants. Nevertheless, Table~\ref{symmetry:table:K_groups} tells us \emph{which} symmetries need to be broken if one wants to find $\Z_2$ topological invariants (only class D does); we expect this to be significant for a first-principles derivation of the bulk-edge correspondence for PhCs. 

\section{The Maxwell-Harper approximation for non-gyrotropic PhCs} 
\label{Maxwell_Harper}
Usually, the frequency bands and Bloch functions are only obtainable numerically for given choices of $\eps$ and $\mu$, and one way to better understand some aspects of light dynamics is to look for simpler model operators which are more amenable to analysis but retain certain features of the full operator. In solid state physics, one such operator is the Harper operator \cite{Hofstadter:Bloch_electron_magnetic_fields:1976}, and the purpose of this section is to motivate a photonic analog. 

Let us consider PhCs of class D or BDI (\ie $C$ is a symmetry of $M_w$). As argued in Sections~\ref{Max_Schroedinger:physical_states} and \ref{symmetry:topological_triviality} as well as \cite[Section~5]{DeNittis_Lein:sapt_photonic_crystals:2013}, if one wants to understand \emph{real} electromagnetic fields, then the \emph{simplest} model for a PhC necessarily includes two symmetrically related bands. Suppose, for instance, that $M_w$ is of class D, then according to Table~\ref{symmetry:table:K_groups} we expect topological effects may still play a role (there are $\Z_2$-invariants) even if the total Chern class of symmetrically chosen bands vanishes.

If the periodic structure is perturbed on the macroscopic level, \ie we replace the periodic material weights $w = (\eps,\mu)$ by $w(\lambda) = (\eps_{\lambda} , \mu_{\lambda})$ where these perturbed material weights 
\begin{align*}
	\eps_{\lambda}(x) &= \tau_{\eps}^{-2}(\lambda x) \, \eps(x) 
	, 
	\\
	\mu_{\lambda}(x) &= \tau_{\mu}^{-2}(\lambda x) \, \mu(x) 
	, 
\end{align*}
are modulated by bounded, strictly positive functions $\tau_{\eps}$ and $\tau_{\mu}$ whose inverses are also bounded. This type of perturbation has been studied theoretically \cite{Raghu_Haldane:quantum_Hall_effect_photonic_crystals:2008,Esposito_Gerace:photonic_crystals_broken_TR_symmetry:2013,DeNittis_Lein:sapt_photonic_crystals:2013} and models effects such as uneven thermal \cite{Duendar_et_al:optothermal_tuning_photonic_crystals:2011,van_Driel_et_al:tunable_2d_photonic_crystals:2004} or uneven strain tuning \cite{Wong_et_al:strain_tunable_photonic_crystals:2004}. 

Quite naturally, the first task in the study of the perturbed Maxwell operator $M_{\lambda} = M_{w(\lambda)}$ is to derive \emph{effective dynamics}, \ie to relate the perturbed to the unperturbed dynamics if one knows something about the initial states. Here, the states one considers are associated to a \emph{relevant family of frequency bands} which is separated by a local gap from the others to prevent band transition. For instance, the bands $\bigl \{ \omega_{n_2}(k) , \omega_{n_{-2}}(k) \bigr \}$ or $\bigl \{ \omega_{n_3}(k) , \omega_{n_4}(k) \bigr \}$. However, we do allow band intersections within the family of relevant bands. 

For the Bloch electron, the analogous situation in quantum mechanics, this is a very well-studied problem \cite{Panati_Spohn_Teufel:sapt_PRL:2002}. One type of effective dynamics are \emph{semiclassical dynamics}: here, the band energy enters the Hamilton function, but the dynamical equations also contain a topological contribution in the form of the Berry curvature which acts as a pseudomagnetic field. Such semiclassical equations of motion, \emph{ray optics equations}, have been proposed for PhCs by Raghu and Haldane  \cite{Raghu_Haldane:quantum_Hall_effect_photonic_crystals:2008} and derived in \cite{Onoda_Murakami_Nagaosa:geometrics_optical_wave-packets:2006,Esposito_Gerace:photonic_crystals_broken_TR_symmetry:2013}. However, Real states can never be supported by single frequency bands so even in the simplest physically relevant situation, one has to work with a conjugate pair of bands $\bigl \{ \omega_{\pm}(k) \bigr \} = \bigl \{ \omega(k) , - \omega(-k) \bigr \}$. Then the single-band effective dynamics needs to be augmented by an analysis of an interband term (\cf the discussion in Section~6 of \cite{DeNittis_Lein:sapt_photonic_crystals:2013}). 

So instead, let us pursue a different, complementary strategy to find effective dynamics in the twin-band case: here, one approximates $\e^{- \ii t M_{\lambda}}$ on the subspace $\ran \Pi_{\lambda}$ defined in terms of the \emph{superadiabatic projection} $\Pi_{\lambda} = \Pi_0 + \order(\lambda)$ \cite{Nenciu:effective_dynamics_Bloch:1991,Panati_Spohn_Teufel:sapt_PRL:2002}. The leading-order term $\Pi_0$ is unitarily equivalent to the family of the projections $\sum_{j = \pm} \sopro{\varphi_j(k)}{\varphi_j(k)}$ onto the eigenspaces of $\omega_{\pm}$. Effective dynamics now means that there exists an effective Maxwell operator $M_{\mathrm{eff}}$ and a unitary $U_{\lambda}$ such that 
\begin{align*}
	\Bigl ( \e^{- \ii t M_{\lambda}} - U_{\lambda}^* \; \e^{- \ii t M_{\mathrm{eff}}} \; U_{\lambda} \Bigr ) \; \Pi_{\lambda} = \order(\lambda^n)
\end{align*}
holds for some $n$. This scheme has recently been implemented rigorously for photonic crystals \cite{DeNittis_Lein:sapt_photonic_crystals:2013}, and $U_{\lambda}$, $\Pi_{\lambda}$ and $M_{\mathrm{eff}}$ have been constructed order-by-order in $\lambda$ via explicit recursion relations. The role of the unitary $U_{\lambda}$ is to map the problem onto a simpler reference Hilbert space which in this case is $L^2(\mathbb{B}) \otimes \C^N$ where $\mathbb{B}$ is the Brillouin zone and in our case $N = 2$ since we are dealing with two non-degenerate bands. 

The leading-order of $M_{\mathrm{eff}}$ has been computed in \cite{DeNittis_Lein:sapt_photonic_crystals:2013} for Bloch bands which individually carry zero Chern charge, and $M_{\mathrm{eff}}$ is the “quantization” of 
\begin{align*}
	\mathcal{M}_{\mathrm{eff}}(r,k) = \tau(r) \, \left (
	\begin{matrix}
		\omega(k) & 0 \\
		0 & - \omega(-k) \\
	\end{matrix}
	\right )
	. 
\end{align*}
This is a matrix-valued function depending on macroscopic position $r$ and crystal momentum $k$, and involves the perturbation via the function $\tau(r) = \tau_{\eps}(r) \, \tau_{\mu}(r)$. After replacing $r$ with $\ii \lambda \nabla_k$ and $k$ with multiplication with $k$, the resulting \emph{Maxwell-Harper operator} 
\begin{align}
	M_{\mathrm{eff}} = \frac{\tau(\ii \lambda \nabla_k)}{2} \, \left (
	\begin{matrix}
		\omega(k) & 0 \\
		0 & - \omega(-k) \\
	\end{matrix}
	\right ) + \mbox{h.c.}
	\label{Maxwell_Harper:eqn:Maxwell_Harper}
\end{align}
is the analog of \emph{“Peierls substitution”} for PhCs. 

One way to further analyze this operator is to assume $\tau$ is periodic on the macroscopic level; for instance, one can think of a finite sample of size $L = \bigl \{ L_1 \, e_1 , L_2 \, e_2 , L_3 \, e_3 \bigr \}$ where the $L_j$ are all positive, large integers, and we impose periodic boundary conditions. This gives rise to a lattice $\Gamma_L$ and a dual lattice $\Gamma_L^*$, and we can expand the modulation 
\begin{align*}
	\tau(r) = \sum_{\gamma^* \in \Gamma_L^*} \hat{\tau}(\gamma^*) \, \e^{+ \ii \gamma^* \cdot r}
\end{align*}
and the frequency band function 
\begin{align*}
	\omega(k) &= \sum_{\gamma \in \Gamma} \hat{\omega}(\gamma) \, \e^{+ \ii k \cdot \gamma}
\end{align*}
in terms of the Fourier coefficients $\hat{\tau}(\gamma^*)$ and $\hat{\omega}(\gamma)$. The operator $M_{\mathrm{eff}}$ can be expressed algebraically in terms of 
\begin{align*}
	\bigl ( {S}_j \psi \bigr )(k) &= \e^{+ \ii {k}_j} \, \psi(k) 
	,
	\\
	\bigl ( {T}_j \psi \bigr )(k) &= \psi \bigl ( k + \tfrac{\lambda}{L_j} \, e_j^{\ast} \bigr ) 
	.  
\end{align*}
These two unitary operators are shifts in real and reciprocal space which satisfy the following commutation relations: 
\begin{align*}
	 {T}_j \, {S}_n= \e^{\ii \frac{\lambda}{L_n} \delta_{jn}}
	 \, {S}_n \, {T}_j
	 \, ,
	 \qquad 
	 \bigl [ {T}_j , {T}_n \bigr ] = 0 = \bigl [ {S}_j , {S}_n \bigr ]
	 \, ,
	 \qquad 
	 j,n = 1,2,3 
	 . 
\end{align*}
After a Fourier transform which maps $L^2(\T^3)$ to $\ell^2(\Gamma)$, one obtains a multiband tight-binding model from~\eqref{Maxwell_Harper:eqn:Maxwell_Harper} just as in condensed matter physics. Simplifying assumptions on the Fourier coefficients of $\omega$ and $\tau$ then lead to tight-binding models which can be analyzed efficiently and perhaps even explicitly. 

These six operators generate a representation of a six-dimensional non-commutative torus on $L^2(\T^3)$ \cite[Chapter~12]{Varilly_Figueroa_Gracia_Bondia:noncommutative_geometry:2001}. Let us denote the $C^\ast$-algebra generated by ${S}_j$ and ${T}_j$ on $L^2(\T^3)$ with $\mathcal{A}^6(\nicefrac{\lambda}{L})$. We have shown that the effective models for the Maxwell dynamics in the twin bands case can be associated with a (diagonal) representative of the non-commutative torus $\mathcal{A}^6(\nicefrac{\lambda}{L}) \otimes \mathrm{Mat}_{\C}(2)$. This analogy allows us to apply all the well-known techniques for Harper operators to the Maxwell-Harper operator~\eqref{Maxwell_Harper:eqn:Maxwell_Harper}. For instance, one can expect to recover a Hofstadter's butterfly-like spectrum \cite{Hofstadter:Bloch_electron_magnetic_fields:1976} which produces a splitting of the two topologically trivial bands $\omega_{\pm}$ into subbands which can carry a non-trivial topology. We stress that in this case the non-trivial effect is due only to an incommensurability between the perturbation parameter $\lambda$ and the lengths $L_j$ of the macroscopic lattice without any magnetic effect.

The operator~\eqref{Maxwell_Harper:eqn:Maxwell_Harper} is just a particular example of a Maxwell-Harper operator; the fact that it is a \emph{diagonal} element of $\mathcal{A}^6(\nicefrac{\lambda}{L}) \otimes \mathrm{Mat}_{\C}(2)$ can be linked to the topological triviality of $\omega_{\pm}$. However, in PhCs the presence of the PH symmetry does not imply the topological triviality of single bands, and the Bloch functions $\varphi_{\pm}(k)$ cannot be used to smoothly diagonalize $M_{\mathrm{eff}}$. Instead, our arguments in Section~\ref{symmetry} suggest to use Real and Imaginary part $\psi_{\Re}$ and $\psi_{\Im}$, and generally one obtains 
\begin{align*}
	\mathcal{M}_{\mathrm{eff}}(r,k) = \frac{\tau(r)}{2} \, 
	\left (
	\begin{matrix}
		\omega(k)- \omega(-k) & -\ii \bigl ( \omega(k)+ \omega(-k) \bigr ) \\
		\ii \bigl ( \omega(k)+ \omega(-k) \bigr ) &  \omega(k)- \omega(-k) \\
	\end{matrix}
	\right ) 
\end{align*}
which after the Peierls substitution produces a non diagonal element of  $\mathcal{A}^6(\nicefrac{\lambda}{L}) \otimes \mathrm{Mat}_{\C}(2)$.

\section{On the role of complex fields in gyrotropic PhCs} 
\label{gyrotropic}
Our explanation for the absence of topological effects in non-gyrotropic PhCs hinged on the presence of the particle-hole symmetry and the assumption that the electromagnetic wave was purely real. However, the material weights in gyrotropic media are complex (such as is the case in \cite{Yeh_Chao_Lin:Faraday_effect:1999,Wang_et_al:edge_modes_photonic_crystal:2008,Kriegler_Rill_Linden_Wegener:bianisotropic_photonic_metamaterials:2010,Wu_Levy_Fratello_Merzlikin:gyrotropic_photonic_crystals:2010}, for instance), and the Maxwell equations are coupled PDEs with complex coefficients. So even if the initial states are real, the time-evolved fields acquire a non-zero imaginary part. That means a distinction between real and complex electromagnetic fields is only meaningful for PhCs with $C$-symmetry, and to understand PhCs with broken $C$-symmetry, the significance of truly complex electromagnetic fields needs to be explored. This is purely a problem of physics, because mathematically no fundamental obstacles arise in the analysis. 

To the best of our knowledge this particular problem has seen very little attention in the literature. The best reference we have been able to track down is \cite{Bergmann:gyrotropic_Maxwell:1982} which covers the case of constant permittivity and permeability; its arguments extend readily to the present setting, but the author stops short of a complete physical interpretation of the complex nature of the plane wave solutions he gets. In standard textbooks (\eg \cite{Jackson:electrodynamics:1998}) complex electromagnetic fields are either discussed in the context of systems with friction or as a convenient way to express solutions of the Maxwell equations in terms of complex plane waves rather than $\sin$ and $\cos$. Neither one of these qualifications applies: In systems with friction or amplification, the eigenvalues of $\eps(x)$ and $\mu(x)$ need to have non-zero imaginary parts. As long as the material weights are hermitian, the field energy~\eqref{Max_Schroedinger:eqn:field_energy} is conserved because the Maxwell operator is selfadjoint. 

Essentially, we see two ways to interpret complex electromagnetic waves: 
\begin{enumerate}[(1)]
	\item One takes the real part of the complex wave. 
	\item One accepts the complex nature of the waves and that only real-valued quantities such as field intensities and the Poynting vector are measured in experiment. 
\end{enumerate}
The problem of strategy (1) is the interpretation of the imaginary part: where does the associated field energy go? And more importantly, our arguments from Section~\ref{symmetry} show the \emph{absence of topological effects} for states of the form $\Re (\mathbf{E} , \mathbf{H} \bigr ) = C \Re \bigl (\mathbf{E} , \mathbf{H} \bigr )$. Clearly, this interpretation yields testable hypothesis that are incompatible with experiment (topological effects in PhCs have been observed \cite{Wang_et_al:edge_modes_photonic_crystal:2008}). 

The second interpretation is consistent with experiment, because it allows for topological effects. Electrodynamics in matter is an \emph{effective theory} that is obtained after a suitable coarse graining procedure and holds only for electromagnetic waves whose \emph{in vacuo} wavelength is large compared to the size of the constituents of the PhC (\eg the size of split-ring resonators). So what is really measured in experiment? Observables in this context are \emph{real}-valued functions in the fields such as the \emph{field intensities} $\babs{\mathbf{E}(t,x)}^2$ and $\babs{\mathbf{H}(t,x)}^2$, and the \emph{Poynting vector} 
\begin{align}
	\mathbf{S} = \Re \, \overline{\mathbf{E}} \times \mathbf{H} 
	. 
	\label{gyrotropic:eqn:Poynting}
\end{align}
The definition of this vector stems from energy conservation (\cite[equation~(38)]{Bergmann:gyrotropic_Maxwell:1982}), namely if 
\begin{align*}
	U_{\mathcal{E}}(t,x) = \frac{1}{2} \Bigl ( \mathbf{E}(t,x) \cdot \eps(x) \mathbf{E}(t,x) + \mathbf{H}(t,x) \cdot \mu(x) \mathbf{H}(t,x) \Bigr ) 
\end{align*}
denotes the energy density, then $S$ satisfies the conservation law 
\begin{align*}
	\partial_t U_{\mathcal{E}} + \nabla_x \cdot \mathbf{S} = 0 
	. 
\end{align*}
Lastly, how are gyrotropic media different? After all, our arguments show that there is nothing mathematically wrong with having complex electromagnetic waves if all that counts is the propagation field intensity. 
\bigskip

\paragraph{Acknowledgements} 
\label{acknowledgements}
G.~D{.} acknowledges support by the Alexander von Humboldt Foundation. M.~L{.} is grateful for the financial support from the Fields Institute. Both of the authors would also like to thank Daniel Ueltschi and Robert Seiringer for the invitation to Warwick where this article was finished. Moreover, the authors would like to thank Ling Lu and the referee for their comments which have helped improve the manuscript. 

%
\begin{appendix}
	\section{Tabulated symmetries} 
	\label{appendix:tabulated_symmetries}
	Given that $W = W^*$ and $W^{-1}$ exists, one can express our material weight tensor as 
	\begin{align}
		W^{-1} = \sum_{j = 0}^3 \sigma_j \otimes w_j 
		= \left (
		\begin{matrix}
			w_0 + w_3 & w_1 - \ii \, w_2 \\
			w_1 + \ii \, w_1 & w_0 - w_3 \\
		\end{matrix}
		\right )
		\label{appendix:tabulated_symmetries:eqn:W_decomposition}
	\end{align}
	where $\sigma_0 = \id$ is the identity and $\sigma_1$, $\sigma_2$ and $\sigma_3$ are the Pauli matrices, and the $w_j$ are hermitian $3 \times 3$ matrices. Given that $\Rot = - \sigma_2 \otimes \nabla^{\times}$, we immediately obtain 
	\begin{align*}
		C \, \Rot \, C &= - \Rot 
		\\
		T_j \, \Rot \, T_j &= - \Rot 
		, 
		&&
		j = 1 , 3 
		\\
		T_2 \, \Rot \, T_2 &= + \Rot 
		\\
		J_j \, \Rot \, J_j &= + \Rot 
		, 
		&&
		j = 1 , 3
		\\
		J_2 \, \Rot \, J_2^* &= - \Rot 
		. 
	\end{align*}
	With the exception of $J_2$ for which $J_2^2 = - \id$, all other symmetries are even. 
	Consequently, $U \, M_w \, U^* = \pm M_w$ translates to the conditions $U \, W \, U^* = \pm W$ for $U = C , T_j , J_j$, $j = 1 , 2 , 3$. These can again be computed very efficiently using the algebraic properties of the Pauli matrices. 
	
	When tabulating these symmetries, one first needs to pick one unitary and/or one antiunitary operator; since the classification uses $U_1$, $U_2$ and the product $U_1 \, U_2$ there are many equivalent choices which yield the same classification scheme. For instance, choosing $C$ and $T_2$ is equivalent to choosing $C$ and $J_2 = T_2 \, C$ or $T_2$ and $J_2$. This helps to reduce the size of Tables~\ref{appendix:tabulated_symmetries:table:++_table}--\ref{appendix:tabulated_symmetries:table:+-_table} further. 
	
	Secondly, two signs need to be chosen, \eg the $(+-)$ combination imposes 
	\begin{align*}
		U_1 \, W \, U_1^* &= + W
		,
		\\
		U_2 \, W \, U_2^* &= - W
		,
	\end{align*}
	for two symmetries $U_1 , U_2 = T_j , J_j , C$, $j = 1 , 2 , 3$. Similarly, we define the $(++)$ and $(--)$ combinations.
	
	Lastly, \emph{linear} symmetries which \emph{commute} with the Maxwell operator are irrelevant for the purpose of CAZ classification. For the benefit of the reader not yet familiar with the CAZ classification, we have labelled those symmetries with “[irrel.]” in Table~\ref{appendix:tabulated_symmetries:table:++_table}. The CAZ class is determined by the remaining symmetry (if present); for instance, a Maxwell operator with $(++)$ symmetries $C$ and $T_2$ is of class D just like any other Maxwell operator with $C \, M_w \, C = - M_w$. In the $(+-)$ and $(--)$ tables, though, we have omitted these superfluous symmetries. 
	
	\newpage
	\begin{table}[H]
		\newcolumntype{Q}{>{\centering\arraybackslash\small} m{2.1cm} }
		\renewcommand{\arraystretch}{1.5}
		\centering
		\begin{tabular}{c | Q | Q | Q | Q  }
			\backslashbox{$+$}{$+$}                       & \emph{none}                                                                                        & $T_1$                                                                                                 & $T_2$                                                                                                  & $T_3$                                                                                                \\ \hline
			\parbox[c][1.2cm][c]{0.5cm}{\centering $T_1$} & \parbox[c]{2.1cm}{\centering $w_0 , w_1 \in \C$ \\ $w_2 , w_3 = 0$ \\ {\footnotesize [AIII]}}      &                                                                                                       &                                                                                                        &                                                                                                      \\ \hline
			\parbox[c][1.2cm][c]{0.5cm}{\centering $T_2$} & \parbox[c]{2.1cm}{\centering $w_0 , w_2 \in \C$ \\ $w_1 , w_3 = 0$ \\ {\footnotesize [irrel.]}}      &                                                                                                       &                                                                                                        &                                                                                                      \\ \hline
			\parbox[c][1.2cm][c]{0.5cm}{\centering $T_3$} & \parbox[c]{2.1cm}{\centering $w_0 , w_3 \in \C$ \\ $w_1 , w_2 = 0$ \\ {\footnotesize [AIII]}}      &                                                                                                       &                                                                                                        &                                                                                                      \\ \hline
			\parbox[c][1.2cm][c]{0.5cm}{\centering $C$  } & \parbox[c]{2.1cm}{\centering $w_0 , w_1 , w_3 \in \R$ \\ $w_2 \in \ii \R$ \\ {\footnotesize [D]}}  & \parbox[c]{2.1cm}{\centering $w_0 , w_1 \in \R$ \\ $w_2 , w_3 = 0$ \\ {\footnotesize [BDI]}}          & \parbox[c]{2.1cm}{\centering $w_0 \in \R,w_2 \in \ii \R$ \\ $w_1 , w_3 = 0$ \\ {\footnotesize [irrel.]}} & \parbox[c]{2.1cm}{\centering $w_0 , w_3 \in \R$ \\ $w_1 , w_2 = 0$ \\ {\footnotesize [BDI]}}         \\ \hline
			\parbox[c][1.2cm][c]{0.5cm}{\centering $J_1$} & \parbox[c]{2.1cm}{\centering $w_0 , w_1 , w_2 \in \R$ \\ $w_3 \in \ii \R$ \\ {\footnotesize [AI]}} & \parbox[c]{2.1cm}{\centering $w_0 , w_1 \in \R$ \\ $w_2 , w_3 = 0$ \\ {\footnotesize [BDI]}}          & \parbox[c]{2.1cm}{\centering $w_0 , w_2 \in \R$ \\ $w_1 , w_3 = 0$ \\ {\footnotesize [irrel.]}}          & \parbox[c]{2.1cm}{\centering $w_0 \in \R,w_3 \in \ii \R$ \\ $w_1 , w_2 = 0$ \\ {\footnotesize [CI]}} \\ \hline
			\parbox[c][1.2cm][c]{0.5cm}{\centering $J_2$} & \parbox[c]{2.1cm}{\centering $w_0 \in \R$ \\ $w_1 , w_2 , w_3 \in \ii \R$ \\ {\footnotesize [C]}}  & \parbox[c]{2.1cm}{\centering $w_0 \in \R, w_1 \in \ii \R$ \\ $w_2 , w_3 = 0$ \\ {\footnotesize [CI]}} & \parbox[c]{2.1cm}{\centering $w_0 \in \R,w_2 \in \ii \R$ \\ $w_1 , w_3 = 0$ \\ {\footnotesize [irrel.]}} & \parbox[c]{2.1cm}{\centering $w_0 \in \R,w_3 \in \ii \R$ \\ $w_1 , w_2 = 0$ \\ {\footnotesize [CI]}} \\ \hline
			\parbox[c][1.2cm][c]{0.5cm}{\centering $J_3$} & \parbox[c]{2.1cm}{\centering $w_0 , w_2 , w_3 \in \R$ \\ $w_1 \in \ii \R$ \\ {\footnotesize [AI]}} & \parbox[c]{2.1cm}{\centering $w_0 \in \R,w_1 \in \ii \R$ \\ $w_2 , w_3 = 0$ \\ {\footnotesize [CI]}}  & \parbox[c]{2.1cm}{\centering $w_0 , w_2 \in \R$ \\ $w_1 , w_3 = 0$ \\ {\footnotesize [irrel.]}}          & \parbox[c]{2.1cm}{\centering $w_0 , w_3 \in \R$ \\ $w_1 , w_2 = 0$ \\ {\footnotesize [BDI]}}         \\
		\end{tabular}
		\caption{This table describes all the interesting cases for the CAZ classification in the $(++)$-case, namely when $U_j \, W \, U_j^*=+W$, $j=1,2$. Each cell contains the possible values of the tensor valued entries $w_0,\ldots,w_3$ of the matrix $W$ according to the choice of $U_1$ (horizontal axis) and $U_2$ (vertical axis). Also the CAZ label is displayed. We observe that the symmetry $T_2$ is irrelevant since $T_2 \, W \, T_2=+W$ implies $[T_2 , M_w]=0$ (linear symmetries commuting with the operator $M$ do not enter in the CAZ classification scheme). Hence, the symmetry class is determined by the other symmetries (if present), \eg a Maxwell operator with symmetries $C$ and $T_2$ is of class D. }
		\label{appendix:tabulated_symmetries:table:++_table}
	\end{table}
	%
	
	
	%
	\begin{table}[H]
		\newcolumntype{Q}{>{\centering\arraybackslash\small} m{2.1cm} }
		\renewcommand{\arraystretch}{1.5}
		\centering
		\begin{tabular}{c | Q | Q   }
			\backslashbox{$-$}{$-$}                       & \emph{none}                                                                                        & $T_2$                                                                                                  \\ \hline
			\parbox[c][1.2cm][c]{0.5cm}{\centering $T_2$} & \parbox[c]{2.1cm}{\centering $w_1 , w_3 \in \C$ \\ $w_0 , w_2 = 0$ \\ {\footnotesize [AIII]}}      &                                                                                                        \\ \hline
			\parbox[c][1.2cm][c]{0.5cm}{\centering $C$  } & \parbox[c]{2.1cm}{\centering $w_0 , w_1 , w_3 \in \ii \R$ \\ $w_2 \in \R$ \\ {\footnotesize [AI]}} & \parbox[c]{2.1cm}{\centering $w_1 , w_3 \in \ii\R$ \\ $w_0 , w_2 = 0$ \\ {\footnotesize [CI]}}         \\ \hline
			\parbox[c][1.2cm][c]{0.5cm}{\centering $J_1$} & \parbox[c]{2.1cm}{\centering $w_0 , w_1 , w_2 \in \ii \R$ \\ $w_3 \in \R$ \\ {\footnotesize [D]}}  & \parbox[c]{2.1cm}{\centering $w_1 \in \ii \R, w_3 \in \R$ \\ $w_0 , w_2 = 0$ \\ {\footnotesize [BDI]}} \\ \hline
			\parbox[c][1.2cm][c]{0.5cm}{\centering $J_2$} & \parbox[c]{2.1cm}{\centering $w_0 \in \ii\R$ \\ $w_1 , w_2 , w_3 \in \R$ \\ {\footnotesize [AII]}} & \parbox[c]{2.1cm}{\centering $w_1, w_3 \in \R$ \\ $w_0 , w_2 = 0$ \\ {\footnotesize [DIII]}}           \\ \hline
			\parbox[c][1.2cm][c]{0.5cm}{\centering $J_3$} & \parbox[c]{2.1cm}{\centering $w_0 , w_2 , w_3 \in \ii \R$ \\ $w_1 \in \R$ \\ {\footnotesize [D]}}  & \parbox[c]{2.1cm}{\centering $w_1 \in \R,w_3 \in \ii \R$ \\ $w_0 , w_2 = 0$ \\ {\footnotesize [BDI]}}  \\
		\end{tabular}
		\caption{This table describes all the interesting cases for the CAZ classification in the $(--)$-case, namely when $U_j \, W \, U_j^* = -W$, $j=1,2$. Each cell contains the possible values of the tensor valued entries $w_0,\ldots,w_3$ of the matrix $W$ according to the choice of $U_1$ (horizontal axis) and $U_2$ (vertical axis). Also the CAZ label is displayed. We observe that the symmetries $T_1$ and $T_3$ are absent since $T_j \, W \, T_j = -W$ implies $[T_j , M_w]=0$, $j = 1 , 3$ (linear symmetries commuting with the operator $M$ do not enter in the CAZ classification scheme). }
		\label{appendix:tabulated_symmetries:table:--_table_short}
	\end{table}
	%
	
	
	%
	\begin{table}[H]
		\newcolumntype{Q}{>{\centering\arraybackslash\small} m{2.1cm} }
		\renewcommand{\arraystretch}{1.5}
		\centering
		\begin{tabular}{c | Q | Q | Q | Q | Q}
			\backslashbox{$+$}{$-$}                       & $T_2$                                                                                                 & $C$                                                                                                   & $J_1$                                                                                                   & $J_2$                                                                                                   & $J_3$                                                                                                   \\ \hline
			\parbox[c][1.2cm][c]{0.5cm}{\centering $T_1$} &                                                                                                       & \parbox[c]{2.1cm}{\centering $w_0 , w_1 \in \ii\R$ \\ $ w_2 , w_3 = 0$ \\ {\footnotesize [BDI]}}      & \parbox[c]{2.1cm}{\centering $w_0 , w_1 \in \ii\R$ \\ $ w_2 , w_3 = 0$ \\ {\footnotesize [BDI]}}        & \parbox[c]{2.1cm}{\centering $w_0\in \ii\R , w_1 \in \R$ \\ $ w_2 , w_3 = 0$ \\ {\footnotesize [DIII]}} & \parbox[c]{2.1cm}{\centering $w_0\in \ii\R , w_1 \in \R$ \\ $ w_2 , w_3 = 0$ \\ {\footnotesize [DIII]}} \\ \hline
			\parbox[c][1.2cm][c]{0.5cm}{\centering $T_3$} &                                                                                                       &  \parbox[c]{2.1cm}{\centering $w_0 , w_3 \in \ii\R$ \\ $ w_1 , w_2 = 0$ \\ {\footnotesize [BDI]}}     & \parbox[c]{2.1cm}{\centering $w_0 \in \ii\R, w_3 \in \R$ \\ $ w_1 , w_2 = 0$ \\ {\footnotesize [DIII]}} & \parbox[c]{2.1cm}{\centering $w_0\in \ii\R , w_3 \in \R$ \\ $ w_1 , w_2 = 0$ \\ {\footnotesize [DIII]}} & \parbox[c]{2.1cm}{\centering $w_0 , w_3 \in \ii\R$ \\ $ w_1 , w_2 = 0$ \\ {\footnotesize [BDI]}}        \\ \hline
			\parbox[c][1.2cm][c]{0.5cm}{\centering $C$  } & \parbox[c]{2.1cm}{\centering $w_1 , w_3 \in \R$ \\ $ w_0 , w_2 = 0$ \\ {\footnotesize [DIII]}}        &                                                                                                       & \parbox[c]{2.1cm}{\centering $w_2 \in \ii\R , w_3 \in \R$ \\ $ w_0 , w_1 = 0$ \\ {\footnotesize [D]}}   & \parbox[c]{2.1cm}{\centering $w_1 , w_3 \in \R$ \\ $ w_0 , w_2 = 0$ \\ {\footnotesize [DIII]}}          & \parbox[c]{2.1cm}{\centering $w_1\in\R , w_2 \in \ii\R$ \\ $ w_0 , w_3 = 0$ \\ {\footnotesize [D]}}     \\ \hline
			\parbox[c][1.2cm][c]{0.5cm}{\centering $J_1$} & \parbox[c]{2.1cm}{\centering $w_1\in\R , w_3 \in\ii \R$ \\ $ w_0 , w_2 = 0$ \\ {\footnotesize [BDI]}} & \parbox[c]{2.1cm}{\centering $w_2\in\R , w_3 \in\ii \R$ \\ $ w_0 , w_1 = 0$ \\ {\footnotesize [AI]}}  &                                                                                                         & \parbox[c]{2.1cm}{\centering $w_1,w_2\in\R $ \\ $ w_0 , w_3 = 0$ \\ {\footnotesize [AI+AII]}}           & \parbox[c]{2.1cm}{\centering $w_1\in\R,w_3\in\ii\R $ \\ $ w_0 , w_2 = 0$ \\ {\footnotesize [BDI]}}      \\ \hline
			\parbox[c][1.2cm][c]{0.5cm}{\centering $J_2$} & \parbox[c]{2.1cm}{\centering $w_1, w_3 \in\ii \R$ \\ $ w_0 , w_2 = 0$ \\ {\footnotesize [CI]}}        & \parbox[c]{2.1cm}{\centering $w_1, w_3 \in\ii \R$ \\ $ w_0 , w_2 = 0$ \\ {\footnotesize [CI]}}        & \parbox[c]{2.1cm}{\centering $w_1, w_2 \in\ii \R$ \\ $ w_0 , w_3 = 0$ \\ {\footnotesize [D+C]}}         & \parbox[c]{2.1cm}{\centering $w_2, w_3 \in\ii \R$ \\ $ w_0 , w_1 = 0$ \\ {\footnotesize [D+C]}}         &                                                                                                         \\ \hline
			\parbox[c][1.2cm][c]{0.5cm}{\centering $J_3$} & \parbox[c]{2.1cm}{\centering $w_1\in\ii \R, w_3 \in \R$ \\ $ w_0 , w_2 = 0$ \\ {\footnotesize [BDI]}} & \parbox[c]{2.1cm}{\centering $w_1\in\ii\R , w_2 \in  \R$ \\ $ w_0 , w_3 = 0$ \\ {\footnotesize [AI]}} & \parbox[c]{2.1cm}{\centering $w_1\in\ii \R, w_3 \in \R$ \\ $ w_0 , w_2 = 0$ \\ {\footnotesize [BDI]}}   & \parbox[c]{2.1cm}{\centering $w_2,w_3\in\R $ \\ $ w_1 , w_0 = 0$ \\ {\footnotesize [AI+AII]}}           &                                                                                                         \\
		\end{tabular}
		\caption{This table describes all the interesting cases for the CAZ classification in the $(+-)$-case, namely when $U_1 \, W \, U_1^* = - W$ and $U_2 \, W \, U_2^* = + W$. Each cell contains the possible values of the tensor valued entries $w_0 , \ldots , w_3$ of the matrix $W$ according to the choice of $U_1$ (horizontal axis) and $U_2$ (vertical axis). Also the CAZ label is displayed. We have again omitted superfluous symmetries. }
		\label{appendix:tabulated_symmetries:table:+-_table}
	\end{table}
	%
\end{appendix}
%


\pagebreak
\printbibliography

\end{document}